\documentclass[12pt]{article}
\usepackage{epsfig}
\usepackage{amsmath}
\usepackage{hhline}
\usepackage{amssymb}
\usepackage{times}

\newlength{\dinwidth}
\newlength{\dinmargin}
\begin{document}  

\newcommand{\etal}{{\it et al.}}
\newcommand{\pom}{{I\!\!P}}
\newcommand{\reg}{{I\!\!R}}
\newcommand{\slowpi}{\pi_{\mathit{slow}}}
\newcommand{\gevsq}{\ensuremath{\mathrm{GeV}^2} }

\newcommand {\gapprox}
   {\raisebox{-0.7ex}{$\stackrel {\textstyle>}{\sim}$}}
\newcommand {\lapprox}
   {\raisebox{-0.7ex}{$\stackrel {\textstyle<}{\sim}$}}
\def\gsim{\,\lower.25ex\hbox{$\scriptstyle\sim$}\kern-1.30ex%
\raise 0.55ex\hbox{$\scriptstyle >$}\,}
\def\lsim{\,\lower.25ex\hbox{$\scriptstyle\sim$}\kern-1.30ex%
\raise 0.55ex\hbox{$\scriptstyle <$}\,}
\newcommand{\pomfluxarg}{f_{\pom / p}\,(x_\pom)}
\newcommand{\dsf}{\mbox{$F_2^{D(3)}$}}
\newcommand{\dsfva}{\mbox{$F_2^{D(3)}(\beta,Q^2,x_{I\!\!P})$}}
\newcommand{\dsfvb}{\mbox{$F_2^{D(3)}(\beta,Q^2,x)$}}
\newcommand{\dsfpom}{$F_2^{I\!\!P}$}
\newcommand{\gap}{\stackrel{>}{\sim}}
\newcommand{\lap}{\stackrel{<}{\sim}}
\newcommand{\rb}[1]{\raisebox{1.5ex}[-1.5ex]{#1}}

\newcommand{\fem}{$F_2^{em}$}
\newcommand{\ftc}{$F_2^{c\bar{c}}$}
\newcommand{\ft}{$F_2$ }

\newcommand{\st}{$\star$}
\newcommand{\trm}{m_{\perp}}
\newcommand{\trp}{p_{\perp}}
\newcommand{\trmm}{m_{\perp}^2}
\newcommand{\trpp}{p_{\perp}^2}
\newcommand{\alp}{\alpha_s}
\newcommand{\alps}{\alpha_s}
\newcommand{\sqrts}{$\sqrt{s}$}
\newcommand{\LO}{$O(\alpha_s^0)$}
\newcommand{\Oa}{$O(\alpha_s)$}
\newcommand{\Oaa}{$O(\alpha_s^2)$}
\newcommand{\PT}{p_{\perp}}
\newcommand{\JPSI}{J/\psi}
\newcommand{\shat}{\ensuremath{\hat{s}}}
%
%
\newcommand{\qsq}{\ensuremath{Q^2}}

\newcommand{\dstar}{\ensuremath{D^*}}
\newcommand{\dzero}{\ensuremath{D^0}}
\newcommand{\pislow}{\ensuremath{\pi_{slow}}}
\newcommand{\goldenchannel}{\ensuremath{D^{*\pm}\to D^0+\pi_{slow}^\pm \to K^{\mp}+\pi^\pm+\pi_{slow}^\pm}}
\newcommand{\dstarp}{\ensuremath{D^{*+}}}
\newcommand{\dstarm}{\ensuremath{D^{*-}}}
\newcommand{\dstarpm}{\ensuremath{D^{*\pm}} }

\def\Journal#1#2#3#4{{#1} {\bf #2} (#3) #4}
\def\NCA{Nuovo Cimento}
\def\NIM{Nucl. Instrum. Meth.}
\def\NIMA{{Nucl. Instrum. Meth.} {\bf A}}
\def\NPB{{Nucl. Phys.} {\bf B}}
\def\PLB{{Phys. Lett.} {\bf B}}
\def\PRL{Phys. Rev. Lett.}
\def\PRD{{Phys. Rev.} {\bf D}}
\def\ZPC{{Z. Phys.} {\bf C}}
\def\EPJC{{Eur. Phys. J.} {\bf C}}
\def\CPC {Comp. Phys. Commun. }

\begin{titlepage}

\noindent
\begin{flushleft}
{\tt DESY 09-165    \hfill    ISSN 0418-9833} \\
{\tt October 2009}                  \\
\end{flushleft}
\noindent
\vspace{2cm}
\begin{center}
\begin{Large}
\boldmath{
{\bf Measurement of the \dstarpm Meson Production Cross Section and \ftc\, at High \qsq\, in \textit{ep} Scattering at HERA}}

\vspace{2cm}

H1 Collaboration

\end{Large}
\end{center}

\vspace{2cm}

\begin{abstract}
The inclusive production of $D^{*\pm}(2010)$ mesons in deep-inelastic 
$e^{\pm} p$ scattering is measured in the kinematic region of photon 
virtuality $100 < Q^2 < 1000\,{\rm GeV}^2$ and inelasticity $0.02 < y < 0.7$. 
Single and double differential cross sections for inclusive \dstar\ meson 
production are measured in the visible range defined by $|\eta(D^*)| < 1.5$ 
and $p_{T}(D^*) > 1.5\,{\rm GeV}$. The data were collected by the H1 
experiment during the period from 2004 to 2007 and correspond to an 
integrated luminosity of $351\,\rm pb^{-1}$.
The charm contribution, \ftc, to the proton structure function \ft 
is determined. The measurements are compared with QCD predictions.
\end{abstract}

\vspace{1.5cm}

\begin{center}
Submitted to {\it Phys. Lett. {\bf{B}}}
\end{center}

\vspace{2cm}
\centerline{Dedicated to the memory of our dear friend and  colleague, Beate Naroska}

\end{titlepage}

%
%
%
\begin{flushleft}

F.D.~Aaron$^{5,49}$,           
C.~Alexa$^{5}$,                
K.~Alimujiang$^{11,51}$,       
V.~Andreev$^{25}$,             
B.~Antunovic$^{11}$,           
S.~Backovic$^{30}$,            
A.~Baghdasaryan$^{38}$,        
E.~Barrelet$^{29}$,            
W.~Bartel$^{11}$,              
K.~Begzsuren$^{35}$
% ...

\bigskip{\it
% ...
$ ^{5}$ National Institute for Physics and Nuclear Engineering (NIPNE) ,
          Bucharest, Romania \\
% ...
$ ^{11}$ DESY, Hamburg, Germany \\
% ...
$ ^{25}$ Lebedev Physical Institute, Moscow, Russia$^{ e}$ \\
% ...
$ ^{29}$ LPNHE, Universit\'{e}s Paris VI and VII, CNRS/IN2P3,
          Paris, France \\
 $ ^{30}$ Faculty of Science, University of Montenegro,
          Podgorica, Montenegro$^{ e}$ \\
% ...
$ ^{35}$ Institute of Physics and Technology of the Mongolian
          Academy of Sciences , Ulaanbaatar, Mongolia \\
% ...
$ ^{38}$ Yerevan Physics Institute, Yerevan, Armenia \\
% ...
$ ^{49}$ Also at Faculty of Physics, University of Bucharest,
          Bucharest, Romania \\
% ...
$ ^{51}$ Supported by the Initiative and Networking Fund of the
          Helmholtz Association (HGF) under the contract VH-NG-401. \\

\bigskip
% ...
$ ^e$ Supported by the Deutsche Forschungsgemeinschaft \\
% ...
}
\end{flushleft}
%

\newpage


\section{Introduction}

The measurement of the charm quark production cross section in deep inelastic scattering (DIS) at HERA 
is a powerful means of testing perturbative quantum chromodynamics (QCD). Within this framework, 
a significant contribution to charm production arises from the boson-gluon fusion process which is 
sensitive to the gluon density in the proton. With increasing photon virtuality, $Q^2$, the charm contribution 
to the inclusive $ep$ scattering cross section rises from a few to up to $20\%$. 
Therefore, the treatment of the effects related to the charm quark contribution,
in particular the mass effects, in perturbative QCD calculations is an important issue in 
the determination of parton distribution functions (PDFs). Different schemes to incorporate these 
effects are available. 

Previous measurements were performed by identifying charm quarks via 
$D$ mesons~\cite{D_H1,D_ZEUS} or using variables which are sensitive 
to the lifetime of heavy flavour hadrons~\cite{VTX_H1,VTX_ZEUS}.
This paper presents a measurement of the \dstarpm meson production cross section in the range of 
large photon virtualities $100<Q^2<1000\,{\rm GeV}^2$. The data were collected with 
the H1 detector at HERA during the running period $2004-2007$ when HERA operated with 
$27.6\,{\rm GeV}$ electrons\footnote{In this paper ``electron'' 
is used to denote both electron and positron.} and $920\,{\rm GeV}$ protons colliding at a centre of 
mass energy of $\sqrt{s}=319\,{\rm GeV}$ and correspond to the integrated luminosity of 
$351\,{\rm pb}^{-1}$.
The measured cross sections are compared to  QCD predictions providing an insight 
into the dynamics of \dstarpm meson production at high \qsq. The charm contribution, 
\ftc, to the proton structure function \ft is determined. 

\section{H1 Detector}

A detailed description of the H1 detector can be found elsewhere~\cite{h1det}. 
In the following only detector components relevant to this analysis are discussed. 
A right handed coordinate system is employed with the origin at the position of the
nominal interaction point that has its $z$-axis pointing in the proton beam, or forward,
direction and $x(y)$ pointing in the horizontal (vertical) direction. The pseudorapidity 
is related to the polar angle $\theta$ by $\eta=-\ln {\tan(\theta/2)}$.

Charged particle tracks are reconstructed in the central tracking detector (CTD). It consists 
of two cylindrical central jet drift chambers (CJC) placed concentrically around the beam-line, 
complemented by the silicon vertex detector~\cite{cst}, inside a solenoid with a homogeneous 
magnetic field of $1.16\,{\rm T}$.  The CJCs are separated by a drift chamber which improves the 
$z$-coordinate reconstruction. A multiwire proportional chamber mainly used for 
triggering~\cite{mwpc} is situated inside the inner CJC. The CTD provides a particle momentum 
measurement over the polar angle $15^\circ<\theta<165^\circ$. The trajectories of charged 
particles are measured with a transverse momentum resolution of 
$\sigma(p_T)/p_T \approx 0.002 \, p_T/{\rm GeV} \oplus 0.015$. The interaction 
vertex is reconstructed from CTD tracks.
The Liquid Argon (LAr) calorimeter~\cite{lar} is used to measure the energy and direction of 
electrons, photons and hadrons. It covers the polar angle range $4^\circ<\theta<154^\circ$ 
with full azimuthal acceptance. Electromagnetic shower energies are measured with a 
precision of $\sigma(E)/E = 12\% / \sqrt{E/{\rm GeV}} \oplus 1\%$ and hadronic energies 
with $\sigma(E)/E = 50\% / \sqrt{E/{\rm GeV}} \oplus 2\%$, as determined in test
beam measurements~\cite{h1testbeam}.
In the backward region, energy measurements are provided by a lead/scintillating-fibre (SpaCal) 
calorimeter~\cite{spacal} covering the angular range $155^\circ<\theta<178^\circ$. 
For electrons a relative energy resolution of $\sigma(E)/E = 7\% / \sqrt{E/{\rm GeV}} \oplus 1\%$
is reached, as determined in test beam measurements~\cite{spacaltestbeam}. The SpaCal also provides time-of-flight 
information for trigger purposes. The luminosity is determined from the rate of the Bethe-Heitler 
reaction $ep\rightarrow ep\gamma$, measured using a photon detector located close to the beam pipe at 
$z=-103\,{\rm m}$, in the backward direction. 

\section{Models of Open Charm Production}
\label{simulations}

Open charm production in electron-proton collisions can be described within different schemes.
At energy scales larger than the charm quark mass, calculations can be performed within the zero-mass 
variable-flavour-number scheme (ZMVFNS)~\cite{zmvfns}, where the charm quark is treated as a massless 
parton in the proton. The fixed-flavour-number scheme (FFNS)~\cite{riemersma} applies close to the 
charm production threshold and takes into account heavy quark mass effects. In the latter scheme all 
quark flavours lighter than charm are treated as massless with massive charm being produced 
dynamically via boson-gluon fusion. A consistent treatment of heavy quarks in perturbative QCD over the full 
energy scale range should be provided through the generalised mass variable flavour number 
scheme (GMVFNS)~\cite{thorne_tung}. 

The prediction of open charm production in FFNS at next-to-leading order (NLO) 
uses separate programs to calculate 
inclusive~\cite{riemersma} and exclusive~\cite{hvqdis} (HVQDIS) quantities.
 The momentum densities of the three light quarks and the gluon in the proton are evolved using 
the DGLAP equations~\cite{dglapref}. For the proton structure the FFNS PDF set MRST2004FF3~\cite{mrstff3} 
is used. The charm quark mass is fixed to $m_c=1.43 \, \rm GeV$ in accordance with this PDF set.
The renormalisation and factorisation scales are set to $\mu_r=\mu_f=\mu_0\equiv \sqrt{Q^2+4m_c^2}$. 
The charm fragmentation fraction into \dstarpm mesons is taken as 
$f(c\to D^*)=23.8\pm0.8\%$~\cite{gladilin} 
from the combination of measurements in $e^+e^-$ experiments. 

In the ZMVFNS calculation at NLO~\cite{zmvfns} a charm mass of $1.6 \, \rm GeV$, renormalisation and factorisation 
scales of $\mu_r=\mu_f=\mu_0=\sqrt{Q^2+4m_c^2}$ and the CTEQ6.6M~\cite{cteq6.6} parton densities
are used. The perturbative fragmentation function~\cite{kkks} is evolved to the chosen scale of 
the transverse \dstarpm momentum in the photon-proton rest frame, $p^*_T(D^*)$. 

Events containing charm quarks are generated using the Monte Carlo programs RAPGAP~\cite{rapgap} 
and CASCADE~\cite{cascade} and are passed through a detailed simulation of the detector response 
to determine the acceptance and efficiency and to evaluate the systematic uncertainties associated 
with the measurements.

The RAPGAP program, based on collinear factorisation and DGLAP evolution, is 
used to generate events containing
$c\bar{c}$ pairs via photon-gluon fusion. The leading order (LO) matrix element 
with massive charm quarks is used. Parton showers, based on the DGLAP evolution, 
model the higher order QCD effects. The charm quark mass is set to $1.43 \, \rm GeV$. The proton structure 
is described by the PDF set CTEQ6.5M~\cite{cteq65} and the factorisation and 
renormalisation scales are set to $\mu_r=\mu_f=\mu_0=\sqrt{Q^2+p^2_T}$.

The CASCADE program is based on the $k_T$ factorisation approach. This calculation of 
the photon-gluon fusion matrix element takes into account the charm quark mass as well as the 
virtuality and transverse momentum of the incoming gluon. Gluon radiation from the 
incoming gluon as well as parton showers from the outgoing charm and anti-charm quarks are implemented 
in a manner which includes angular ordering constraints. The gluon density of the proton is evolved according 
to the CCFM equations~\cite{ccfm}. The charm quark mass and the renormalisation scale are set to
$m_c=1.5 \, \rm GeV$ and $\mu_r=\sqrt{Q^2+p^2_T}$, respectively. The unintegrated gluon distribution is 
described by the parametrisation set A0~\cite{a0}. 

The kinematics of \dstarpm production depend not only on the charm quark production but also on the 
$c \to \dstarpm$ fragmentation process. The charm fragmentation function has been measured at 
H1~\cite{h1frag} using inclusive \dstarpm meson production. The Kartvelishvili 
fragmentation function~\cite{kart}, which is controlled by a single parameter $\alpha$, is used. 
The parameter values corresponding to the programs used in the present analysis are shown in 
Table~\ref{tab:fragmentation}. They depend on the centre of mass energy squared of the hard process, \shat.
To obtain the visible \dstarpm production cross sections in HVQDIS, charm quarks are 
fragmented independently in the photon-proton centre of mass frame into \dstarpm mesons 
according to Kartvelishvili function.
In the RAPGAP and CASCADE programs hadronisation is performed using the Lund String Model~\cite{lundf,pythia62}. 
The momentum fraction of the charm quark carried by the \dstarpm meson is modelled according 
to the Bowler parameterisation~\cite{bowler}. The longitudinal part of the fragmentation function 
is reweighted to the Kartvelishvili function. 
\renewcommand{\arraystretch}{1.2}
\begin{table}[htb]
\begin{center}
\begin{tabular}[t]{|l|c|c|} \hline
    Model & $\shat < 70 \,{\rm GeV}^2$ & $\shat > 70\,{\rm GeV}^2$  \\
\hline 
    HVQDIS &  $\alpha=\phantom{1} 6.0^{+1.1}_{-1.3}$ & $\alpha=3.3^{+0.4}_{-0.4}$ \\
    RAPGAP &  $\alpha=10.3^{+1.9}_{-1.6}$ & $\alpha=4.4^{+0.6}_{-0.5}$ \\
    CASCADE & $\alpha=\phantom{1} 8.4^{+1.4}_{-1.1}$ & $\alpha=4.5^{+0.6}_{-0.6}$ \\
\hline
    \end{tabular}
   \caption{Parameter $\alpha$ of the Kartvelishvili fragmentation function 
            as used in the analysis.}
   \label{tab:fragmentation}
\end{center}
\end{table}
\renewcommand{\arraystretch}{1.0}

The contribution of beauty production is estimated using the HVQDIS calculation, with hadronisation 
corrections determined using RAPGAP. The PDF set MRST2004FF3 is used with
$m_b=4.3\,{\rm GeV}$ and 
$\mu_r=\mu_f=\mu_0\equiv \sqrt{Q^2+4m_b^2}$. The fraction of beauty quarks producing \dstarpm 
mesons is taken as $f(b\to D^*)= 17.3\pm 2.0\%$~\cite{beautyfraction}.

\section{Event Selection and Signal Extraction}
\label{event_selection}
DIS events are selected by requiring a compact electromagnetic cluster in either the LAr or SpaCal 
calorimeters, which is taken to be the energy deposit of the scattered electron. 
The cluster has to be associated 
to a track reconstructed in the CTD.
The events are triggered by either a coincidence of a SpaCal cluster and a signal from the CJC, 
or by the presence of a LAr cluster and a signal from the proportional chambers.
The hadronic final 
state (HFS) particles are reconstructed using a combination of tracks and calorimeter deposits 
in an energy flow algorithm~\cite{hadroo2} which avoids double-counting. The event kinematics 
including the photon virtuality \qsq, the Bjorken scaling variable $x$ and the inelasticity 
variable $y$ are reconstructed with the $e\Sigma$ method~\cite{sigmamethod}, which uses the 
scattered electron and the HFS. The measurement is performed in the kinematic region $100<Q^2<1000 \, \rm GeV^2$ and 
$0.02<y<0.7$.

The \dstarpm mesons from the decays $D^{*\pm}(2010)\rightarrow D^0(1865)\pi_{slow}^{\pm}\rightarrow(K^\mp\pi^\pm)\pi_{slow}^\pm$ are reconstructed using the tracks in the CTD. 
The branching ratio for this channel amounts to $B=2.63\pm 0.04\%$ \cite{PDG08}.
The invariant mass of the $K\pi$ combination is required to satisfy $|m(K\pi)-m(D^0)|<80 \, \rm MeV$ where 
$m(D^0)=1864.84\,{\rm MeV}$~\cite{PDG08}. The decay angle $\theta^*$ of the kaon in the rest frame of 
the $D^0$ is restricted to $\cos\theta^*>-0.7$, in order to reduce the background, 
which strongly increases towards $\cos (\theta^*)=-1$ as opposed to the $D^0$, which decays isotropically. 
To further reduce the combinatorial background, a $Q^2$-dependent 
cut on the \dstarpm transverse momentum, $p_T(D^*)/ \rm GeV>(3\cdot[\log(Q^2/ \rm GeV^2)-2]+2)$, 
is applied. This criterion accounts for the increasing transverse momentum of the hadronic 
final state with rising \qsq. 

\begin{figure}[htb]
\centering
\includegraphics*[width=6.5cm]{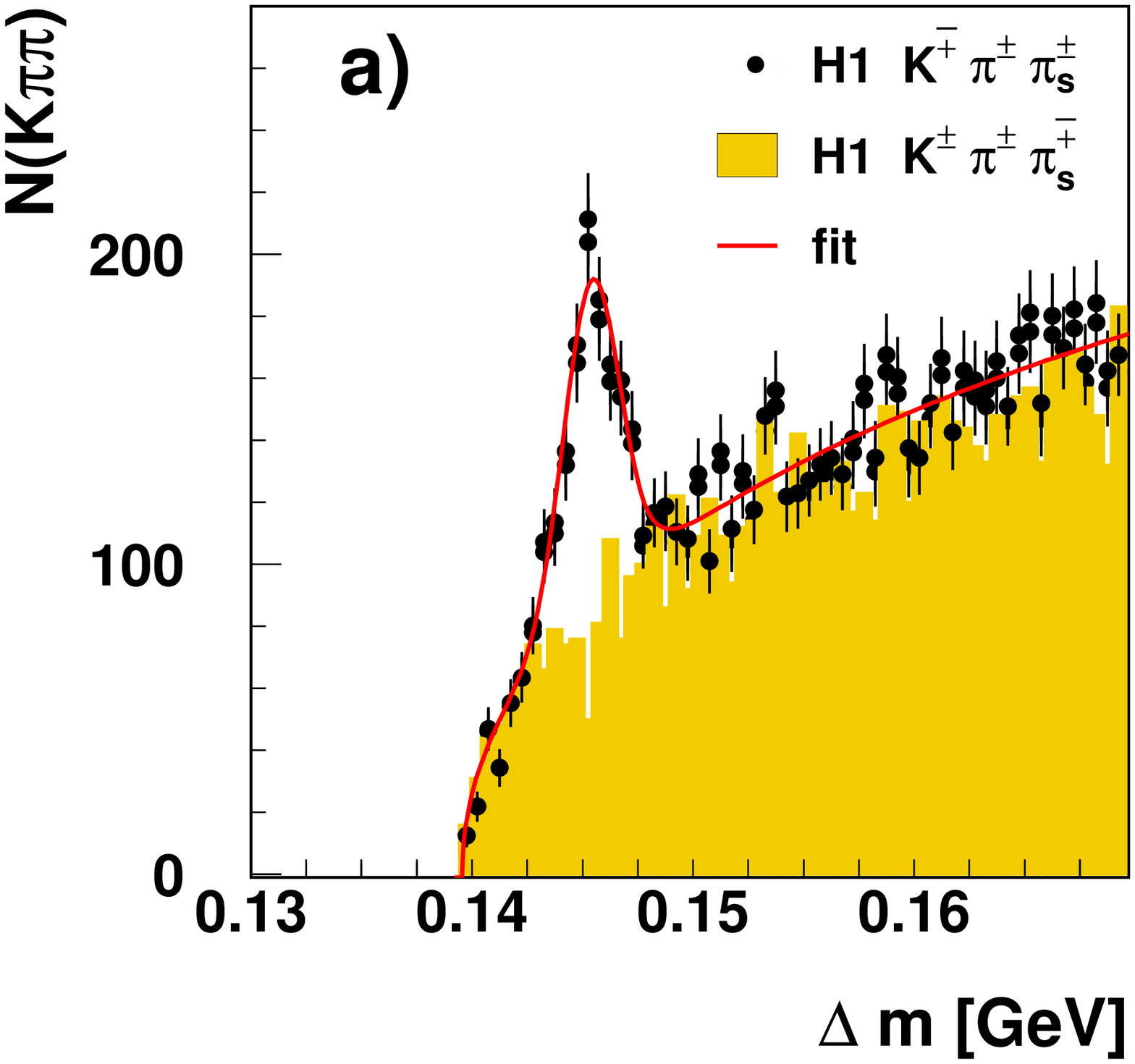} 
\includegraphics*[width=6.5cm]{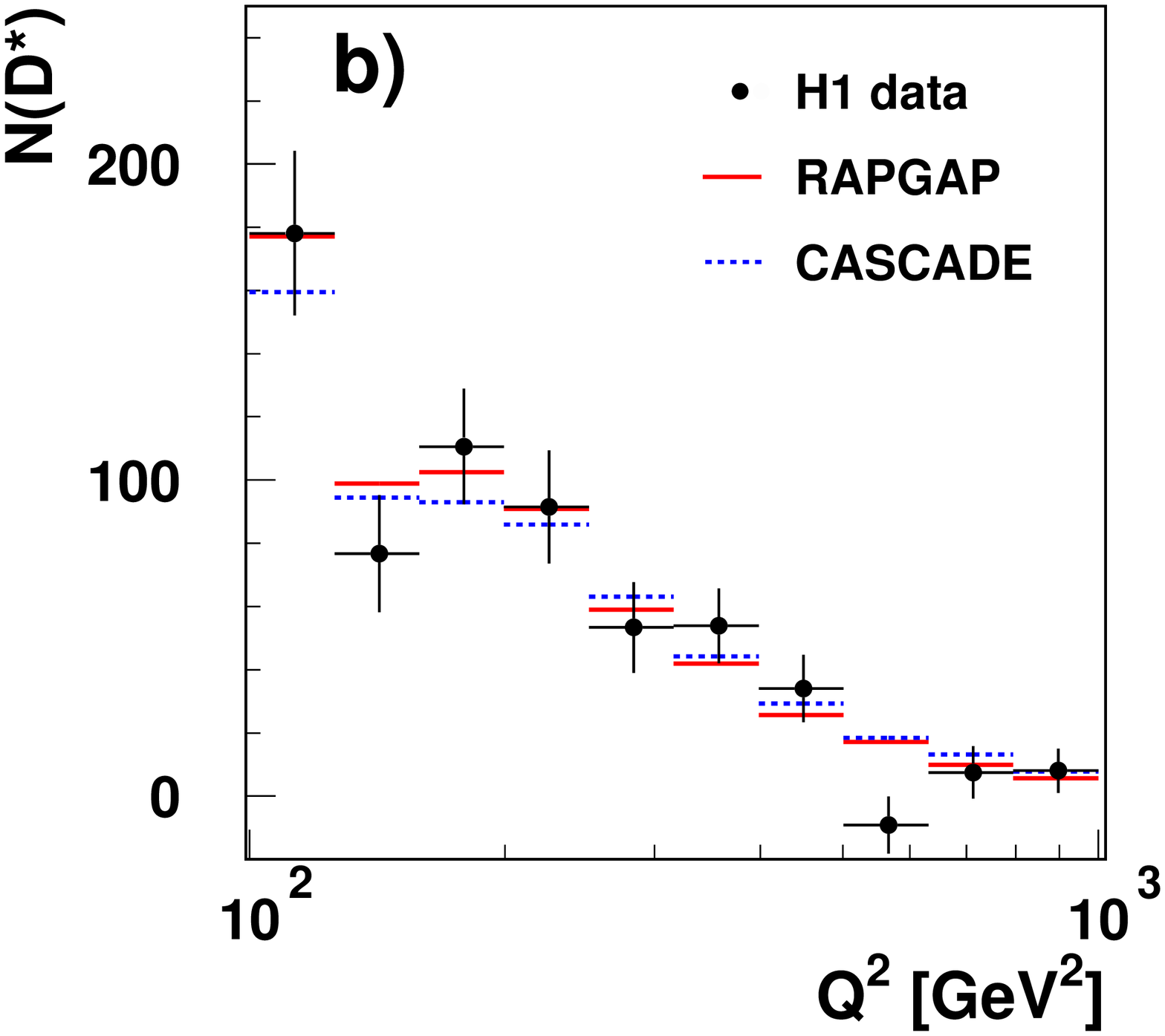}
\includegraphics*[width=6.5cm]{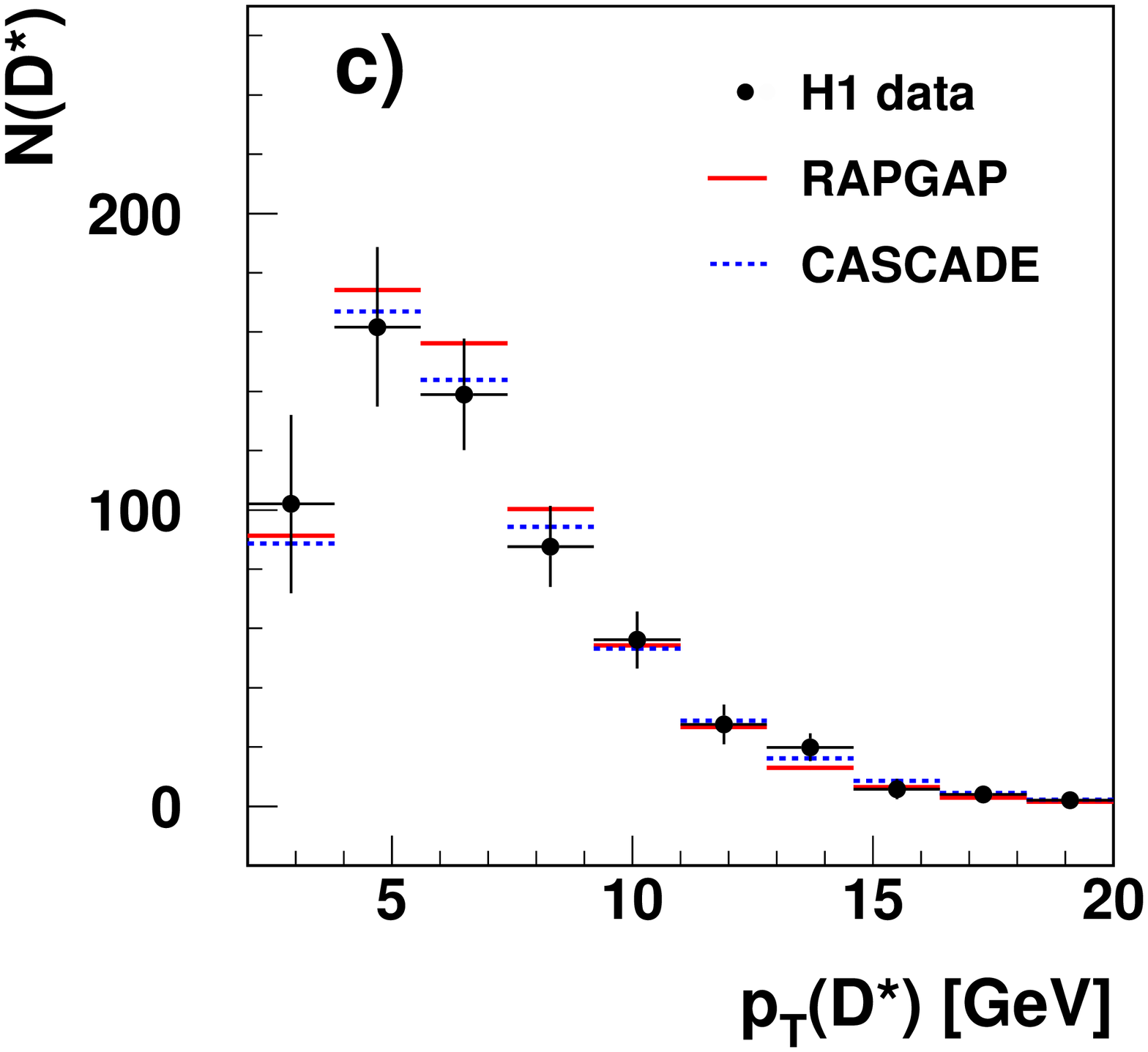} 
\includegraphics*[width=6.5cm]{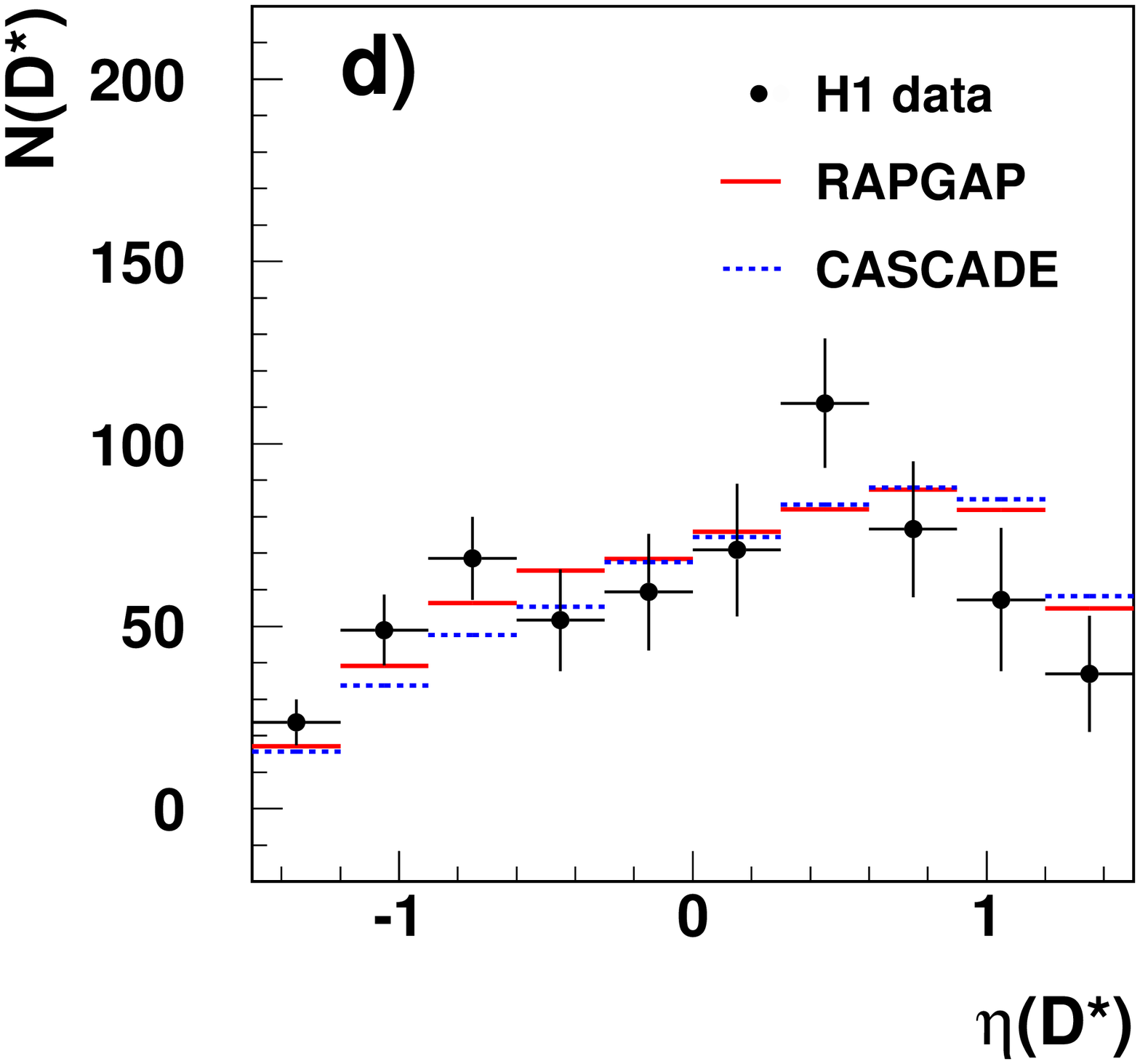}
\caption{a) Distribution of $\Delta m=m(K\pi\pi)-m(K\pi)$ for \dstarpm candidates ($K^\mp \pi^\pm \pi_s^\pm$) and 
for wrong charge combinations ($K^\pm \pi^\pm \pi_s^\mp$) in the accepted $D^0$ mass window. The fit
function is also shown.  Comparisons at the detector level between the \dstarpm data 
sample and the reweighted Monte Carlo models are presented. 
Background-subtracted distributions are shown as a function of $Q^2$ (b),
$p_T(D^*)$ (c) and $\eta(D^*)$ (d).}
\label{fig:signal}
\end{figure}

The \dstarpm candidates in the pseudorapidity range $|\eta(D^*)|<1.5$ are
selected using the mass difference method~\cite{deltam}.
In Fig.~\ref{fig:signal}(a) the distribution of the mass difference $\Delta m=m(K\pi\pi)-m(K\pi)$ is shown for 
the selected data sample. A clear peak is observed around the nominal mass difference of $145.4 \, \rm MeV$. 
Wrong charge $K^{\pm}\pi^\pm \pi^{\mp}$ combinations with $K^\pm \pi^\pm$ pairs in the accepted $D^0$ mass range
are used to describe the combinatorial background. 

The number of \dstarpm mesons is determined in each analysis bin from a simultaneous fit 
to the signal and the background distributions. The Crystal Ball function~\cite{CB} is used 
for the signal description and the Granet parametrisation~\cite{granet} for the background. 
Several fit parameters in the single and double-differential distributions are fixed using 
the full data sample and the Monte Carlo predictions~\cite{brinkmann}.

The cross section presented in this paper corresponds to the kinematic range
summarised in Table~\ref{tab:range}.
The $p_T(D^*)$ and $\eta(D^*)$ range is chosen to be the same as in previous H1 analyses~\cite{D_H1} at 
lower \qsq. The Monte Carlo simulation is used for the extrapolation down to $p_T(D^*)=1.5 \, \rm GeV$. 
This extrapolation typically leads to a $15\%$ increase in the cross section. With all the selection 
cuts, the average acceptance amounts to around $30\%$.
 
\renewcommand{\arraystretch}{1.2}
\begin{table}[htb]
\begin{center}
\begin{tabular}[t]{|ll|} \hline
    Photon virtuality $Q^2$ & $100<Q^2<1000\,{\rm GeV}^2$  \\ 
    Inelasticity $y$ & $0.02<y<0.7$  \\ 
    Pseudorapidity of \dstarpm & $-1.5<\eta(D^*)<1.5$  \\ 
    Transverse momentum of \dstarpm  & $p_{T}(D^*)>1.5\,{\rm GeV}$  \\ \hline
    \end{tabular}
   \caption{Definition of the kinematic range of the present analysis.}
   \label{tab:range}
\end{center}
\end{table}
\renewcommand{\arraystretch}{1.0}

The inclusive \dstarpm production cross section is studied differentially in the kinematic variables 
\qsq, $x$, $p_T(D^*)$, $\eta(D^*)$ and the \dstarpm inelasticity 
$z(D^*)$, which corresponds to the fraction of the virtual photon momentum carried by the \dstarpm meson. 
The \dstarpm inelasticity is determined as $z(D^*)=P\cdot p_{D^*}/P\cdot q =(E-p_z)_{D^*}/2yE_e$, 
where $E_e$ is the energy of the incoming electron and $P$, $q$ and $p_{D^*}$ denote the four-momenta 
of the incoming proton, the exchanged photon and the \dstarpm meson, respectively. 
The cross section for \dstarpm meson production is calculated from the observed number 
of \dstarpm candidates $N_{D^{*\pm}}$, according to: 
\begin{equation}
\sigma_{\rm{vis}}(e^+p \rightarrow e^+D^{*\pm} X)=
\frac{N_{D^{*\pm}}~\cdot (1-r)}
{ {\cal L}_{int} \cdot B \cdot\epsilon\cdot(1+\delta_{rad})} \ \ ,
\label{sigman}
\end{equation}
where $\epsilon$ is the reconstruction efficiency, $r$ the contribution from reflections, 
${\cal L}_{int}$ the integrated luminosity, $B$ the branching ratio and $\delta_{rad}$ 
denotes the radiative corrections. 

The reconstruction efficiency accounts for the trigger efficiency and the
detector acceptance 
and is determined using the Monte Carlo simulation. 
For this purpose charm DIS events are generated using both the RAPGAP and CASCADE 
programs and the average efficiency is used.
For the efficiency determination, RAPGAP is reweighted in $Q^2$ and CASCADE is reweighted in 
$p_T(D^*)$ in order to optimise the data description. The kinematic distributions of 
the \dstarpm candidates compared with the reweighted Monte Carlo predictions are 
shown in Fig.~\ref{fig:signal}(b)-(d).

The contribution $r$ of reflections in the $D^0$ mass window from $D^0$ decay channels 
other than that considered in this analysis is estimated using the Monte Carlo simulation. 
This contribution amounts to $r=(4.4\pm 0.5)\%$ independently of the \dstarpm transverse momentum. 
The radiative corrections $\delta_{rad}$ are determined using RAPGAP interfaced 
to HERACLES~4.1~\cite{heracles} and amount to $3\%$ on average.
The photoproduction background estimated using data~\cite{brinkmann} is not subtracted, but does not exceed $2.7\%$. 
The fraction of \dstarpm mesons originating from $b\bar{b} $ events is estimated as described in 
section~\ref{simulations}. 
It amounts to $4\%$ on average and is included by definition in the inclusive \dstarpm cross section. 
However, for the extraction of \ftc, the predicted contribution from $b\bar{b}$ production 
is subtracted from the data.

\section{Systematic Uncertainties}
\label{uncertainties}
The systematic uncertainties are estimated by varying the input parameters to the Monte Carlo 
simulations within the experimental precision at the reconstructed level or the range 
allowed by the theoretical models at the generator level. 
The following correlated uncertainties are taken into account:
\begin{itemize}
\item The uncertainty on the hadronic energy scale is propagated to the measurement by changing the
hadronic energy by $\pm2\%$ ($\pm3\%$) for events where the scattered electron
is detected in the LAr (SpaCal) calorimeter. 
The uncertainty due to the scattered electron measurement is estimated by varying the 
electron energy by $\pm1\%$ and the polar angle by $\pm3\,{\rm mrad}$, respectively.

\item The trigger efficiency, luminosity and $D^*\rightarrow K\pi\pi$ branching 
ratio are known with uncertainties of $1\%$, $3.2\%$ and $1.5\%$, respectively. 
An uncertainty of $1.2\%$ on the cross-section measurement arises due to the uncertainty 
on the photoproduction background.

\item The uncertainty on the reconstruction efficiency is taken as half of the difference between the
two simulations, RAPGAP and CASCADE. This also covers the uncertainty on the extrapolation to 
$p_T(D^*)=1.5 \, \rm GeV$. The uncertainty in the efficiency determination due to the charm 
fragmentation model is estimated by varying the Kartvelishvili parameter $\alpha$ within its error as described 
in section~\ref{simulations}. The uncertainty due to the choice of PDFs is estimated by using the 
CTEQ6L(LO)~\cite{cteq6l} parton densities in RAPGAP and the A2 set~\cite{a2} in CASCADE as 
alternatives.
\end{itemize}
\noindent
The following uncorrelated systematic uncertainties are accounted for:
\begin{itemize}
\item The signal shape and the invariant mass resolutions of the data are not fully reproduced 
by the Monte Carlo simulation. The errors on the \dstarpm signal extraction are 
determined by varying the fit parameters within their uncertainties.  
The fraction of events outside the \dzero\ mass 
window is determined using the Monte Carlo simulation. Half of this fraction is taken 
as a systematic error to account for the uncertainty on the \dzero\ mass resolution.

\item An uncertainty of $0.5\%$ is assigned to the contribution from reflections to account for a
possible $ p_T $ dependence. The uncertainty of the QED radiative corrections is
$1.5\%$.
\end{itemize}
\noindent
The following uncertainties are treated as partly correlated:
The charged particle reconstruction uncertainty of $2.17\%$, which translates
to $6.5\%$ per $D^{*\pm}$ and the uncertainty on the electron track-cluster 
matching of $2\%$. The above uncertainties are added 
in quadrature to derive the experimental systematic error.

The theoretical uncertainties on the HVQDIS prediction are estimated by varying
the input parameters as follows. The charm mass is varied from $1.3$ to $1.6 \, \rm GeV$. 
The factorisation and renormalisation scales $\mu_f=\mu_r$ are varied simultaneously 
from $0.5\mu_0$ to $2\mu_0$. The fragmentation parameter is varied within its error 
as described in section~\ref{simulations}. The parton density set CTEQ5F3~\cite{cteq5f3} 
is used as an alternative to MRST2004FF3. The resulting uncertainties, together with 
the error on $f(c \to D^*)$, are added in quadrature and are correlated between the bins.
The uncertainties on  the ZMFVNS prediction~\cite{zmvfns} are estimated 
by variation of the renormalisation and factorisation scales simultaneously 
from $0.5\mu_0$ to $2\mu_0$. 

\section{\boldmath{\dstarpm} Production Cross Section}

The total inclusive cross section for \dstarpm production in the phase space covered in this analysis 
(Table~\ref{tab:range}) is measured to be:
$$\sigma_{vis}(e^+p \rightarrow e^+D^{*\pm} X)=225\pm 14({\rm stat.})\pm 27({\rm
syst.})~\rm{pb}\ \ .$$ 

\begin{figure}[htb]
\begin{center}
\includegraphics*[width=6.5cm,height=6.cm]{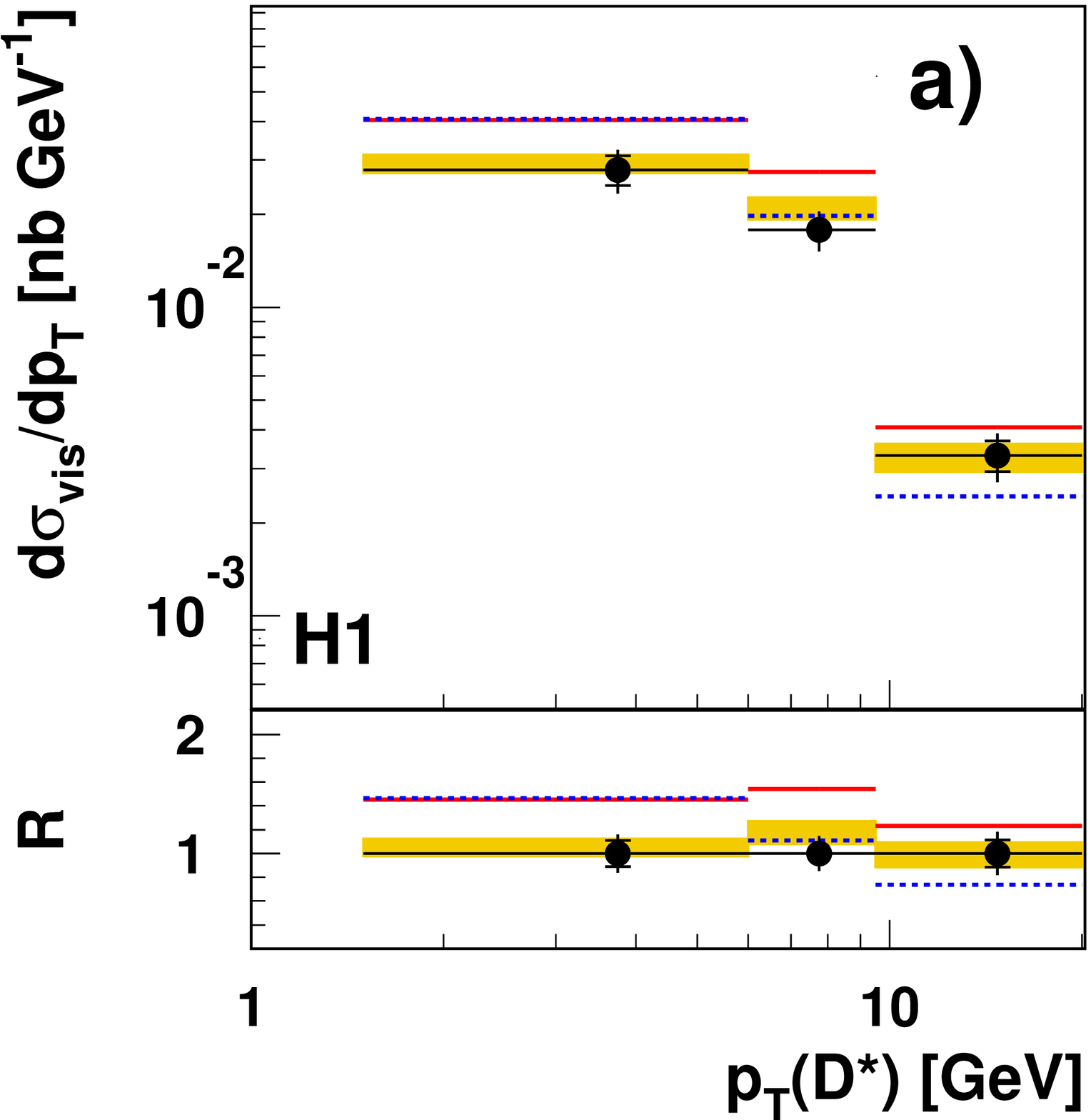}
\includegraphics*[width=6.5cm,height=6.cm]{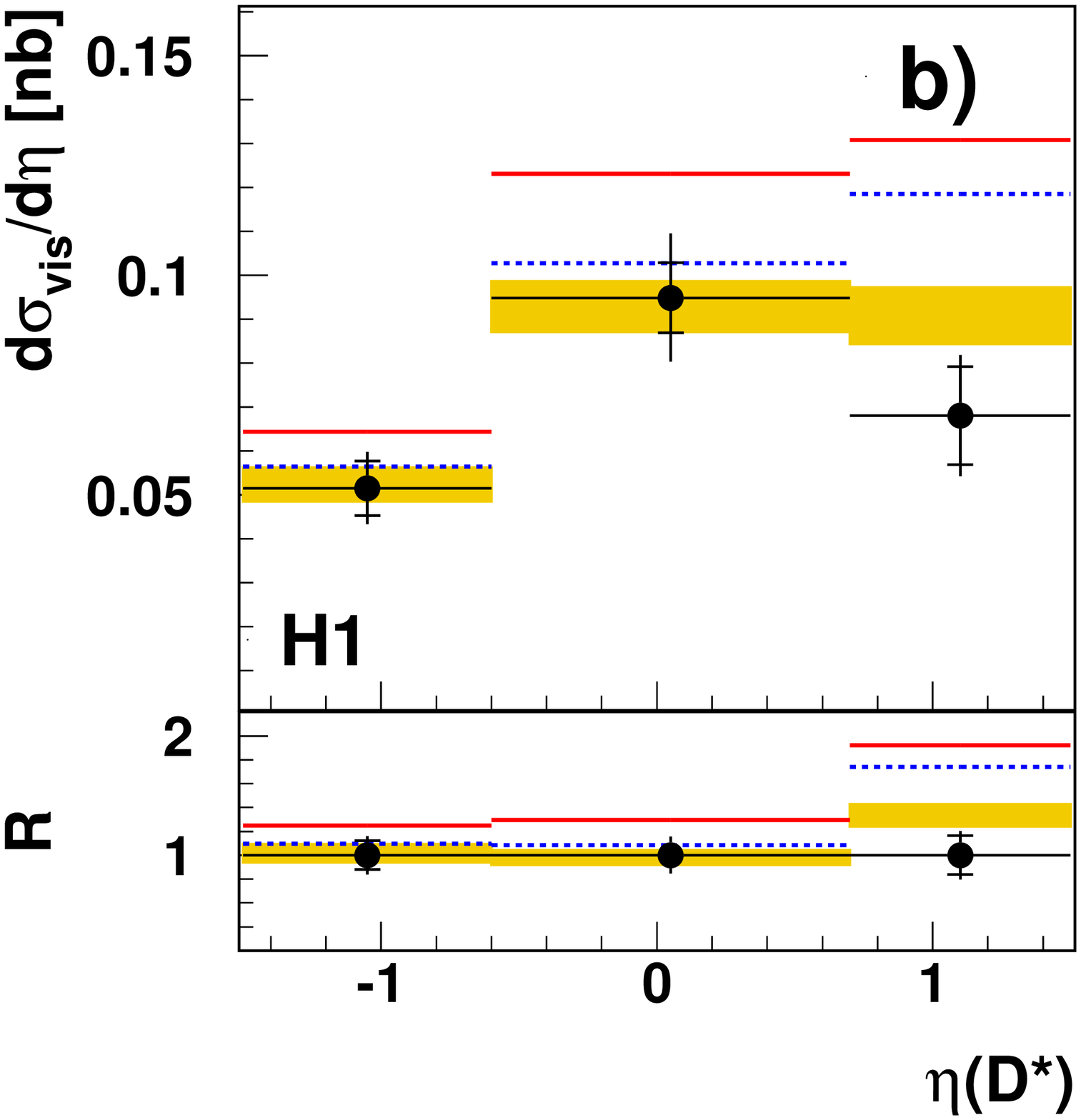}
\includegraphics*[width=6.5cm,height=6.cm]{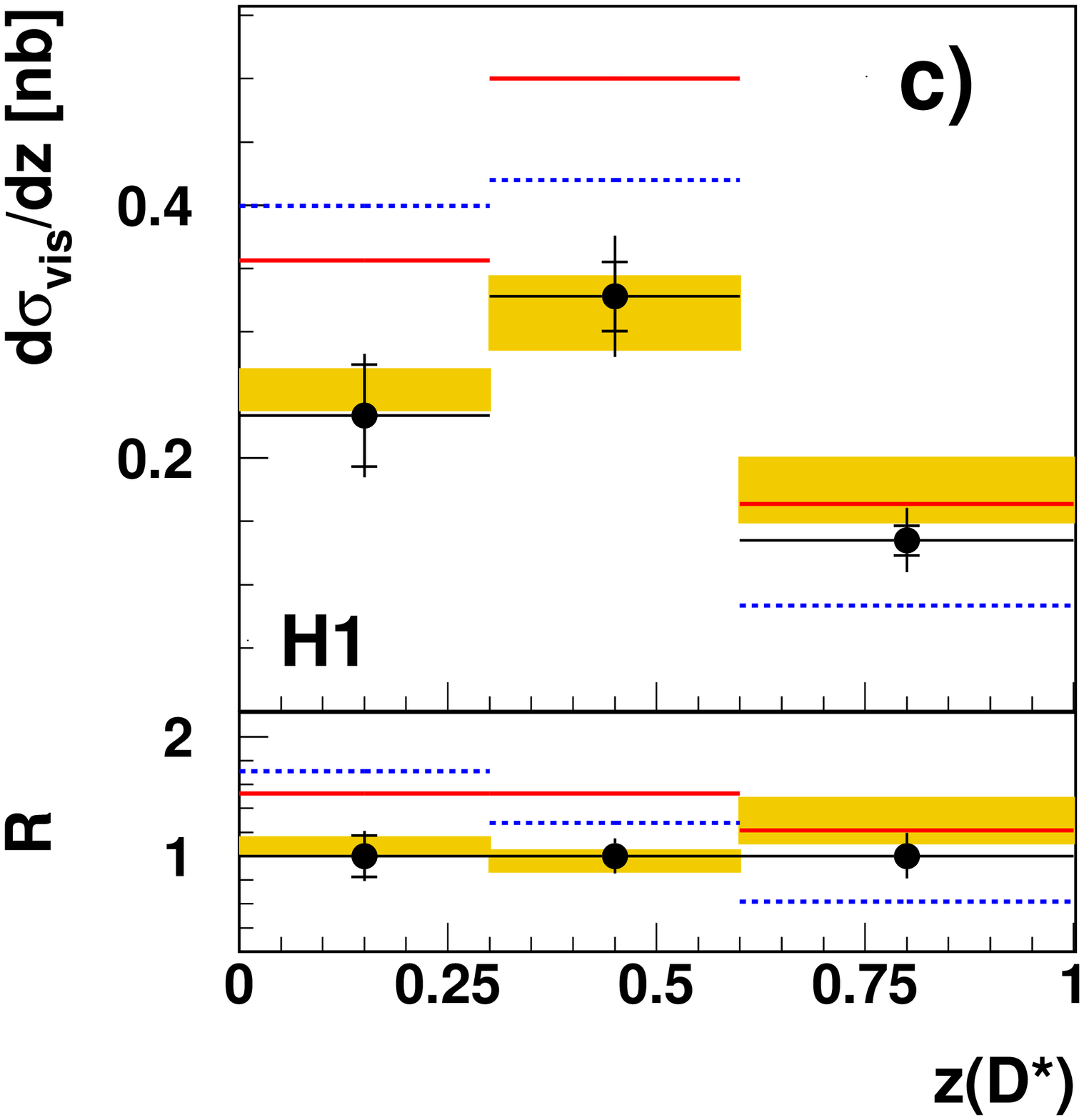}
\includegraphics*[width=6.5cm,height=6.cm]{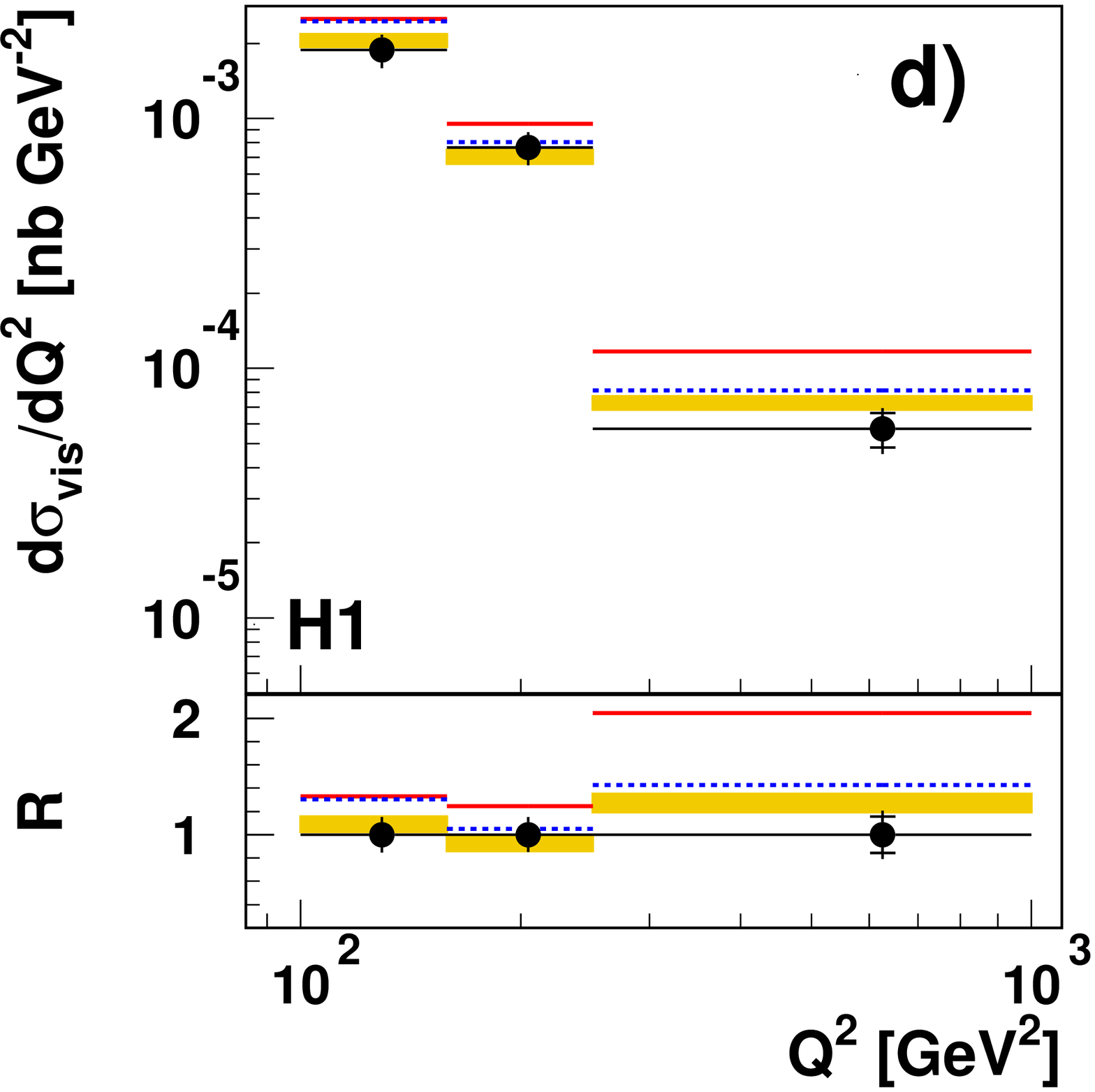}
\includegraphics*[width=6.5cm,height=6.cm]{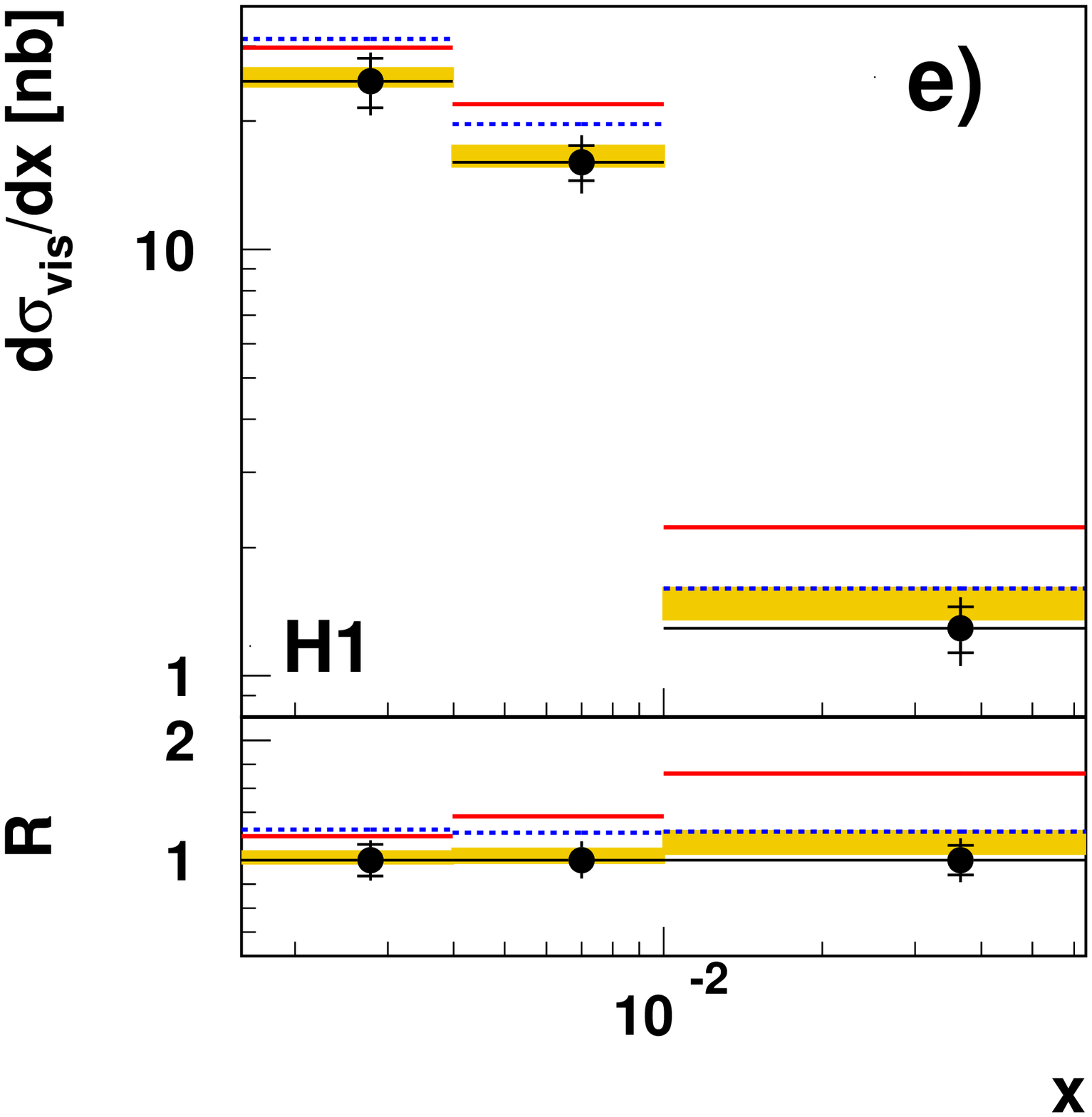}
\includegraphics*[width=6.5cm,height=6.cm]{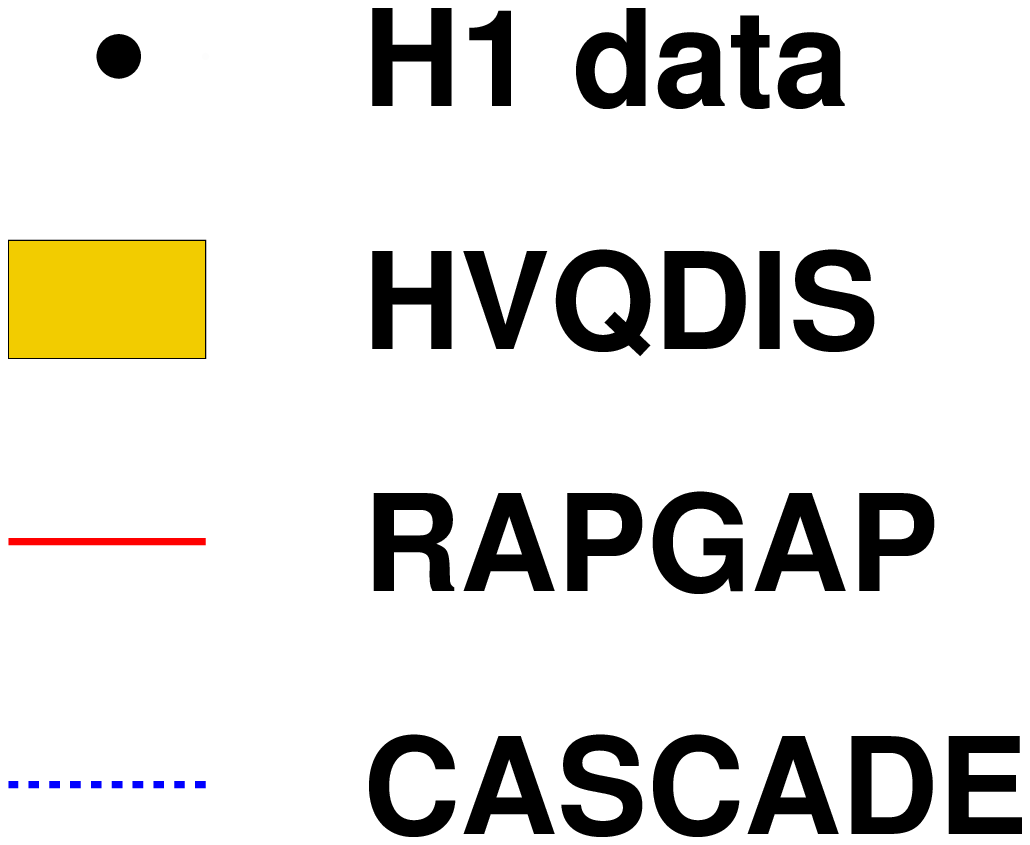}
\end{center}
\caption{Differential cross sections for inclusive \dstarpm meson production as a function of 
$p_T(D^*)$, $\eta (D^*)$, $z(D^*)$, $Q^2$ and $x$. The inner error bars indicate the 
statistical uncertainties, the outer error bars show the statistical and systematic uncertainties added 
in quadrature. The expectations of CASCADE (dashed line) and RAPGAP (solid line) are obtained 
using the parameters as described in section~\ref{simulations}. The band of the HVQDIS prediction 
(shaded) is obtained using the parameter variation described in section~\ref{uncertainties}. 
The ratio $R=\sigma_{theory}/\sigma_{data}$ is also shown. In the case of HVQDIS the theoretical 
uncertainties are taken into account. The inner error bars on the data points at $R=1$ display 
the relative statistical errors, and the outer error bars show the relative statistical and 
systematic uncertainties added in quadrature.}
\label{fig:dsigma_xq}
\end{figure}

The corresponding predictions from RAPGAP, CASCADE and HVQDIS amount to
$322\,{\rm pb}$, $279\,{\rm pb}$, 
and $241^{+14}_{-15}\,{\rm pb}$, respectively, including the $b\bar{b}$ contribution.
In Fig.~\ref{fig:dsigma_xq} and Table~\ref{Tab:dsigma} differential cross 
sections are presented as a function of the DIS kinematic variables $x$ 
and $Q^2$ and as a function of the $D^*$ variables $p_T(D^*)$, $\eta(D^*)$ and $z(D^*)$ . 
The data are compared to the expectations from the HVQDIS calculation and from
the RAPGAP and CASCADE 
Monte Carlo simulations. Neither Monte Carlo simulation describes the shape and normalisation 
of the \dstarpm kinematic distributions well, in contrast to the measurement~\cite{D_H1} at 
lower \qsq. The HVQDIS calculation agrees with the data within the theoretical uncertainties. 

\begin{figure}[htb]
\begin{center}
\includegraphics*[width=12.5cm]{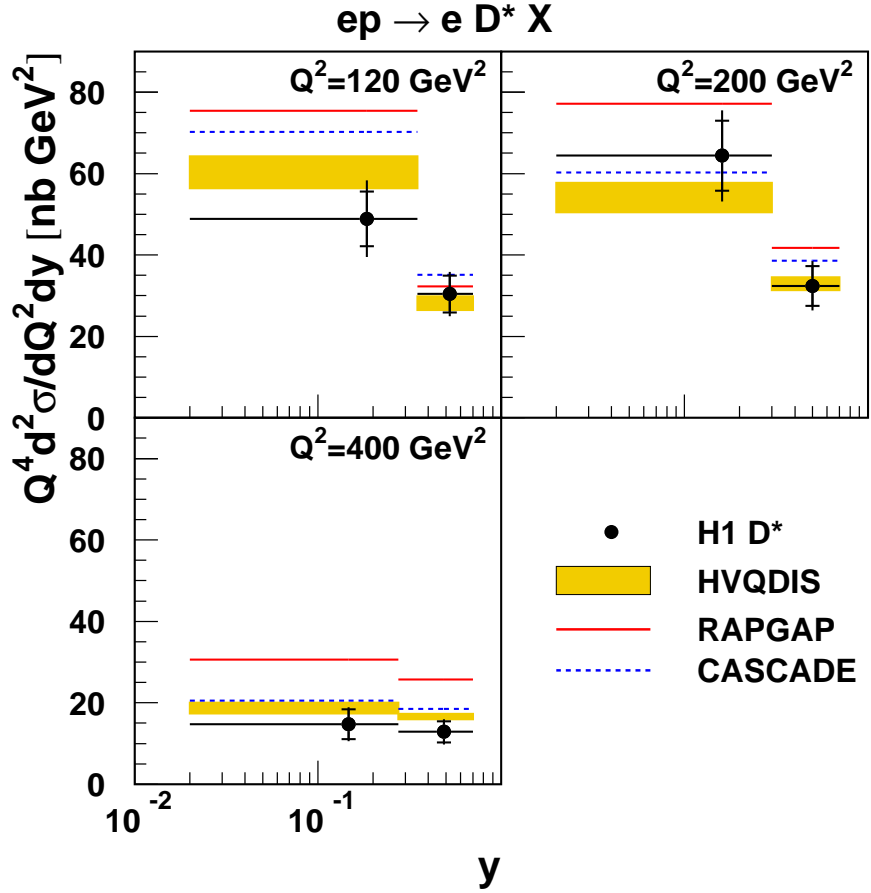}
\end{center}
\caption{Double-differential cross sections for \dstarpm production as a function of $y$ in different $Q^2$ bins.
For the purpose of presentation the cross sections are multiplied by $Q^4$. The data (closed symbols) are 
shown with the statistical (inner error bars) and total (full error bars) uncertainties.
Predictions from the RAPGAP (solid line) and CASCADE (dashed line) Monte Carlo simulations 
and the HVQDIS NLO calculation (shaded area) are also shown.}
\label{fig:d2sigma_dydq2}
\end{figure}

In Fig.~\ref{fig:d2sigma_dydq2} and Table~\ref{Tab:d2sigma} the double differential 
cross sections are shown as a function of $y$ for different bins in $Q^2$. The data are compared to 
the expectations of the HVQDIS calculation as well as to the RAPGAP and CASCADE simulations. HVQDIS 
describes the data well. Except for the first $(Q^2,y)$ bin, the same holds for CASCADE. RAPGAP significantly 
overestimates the visible cross section.  

\begin{figure}[htb]
\begin{center}
\includegraphics*[width=6.5cm,height=6.4cm]{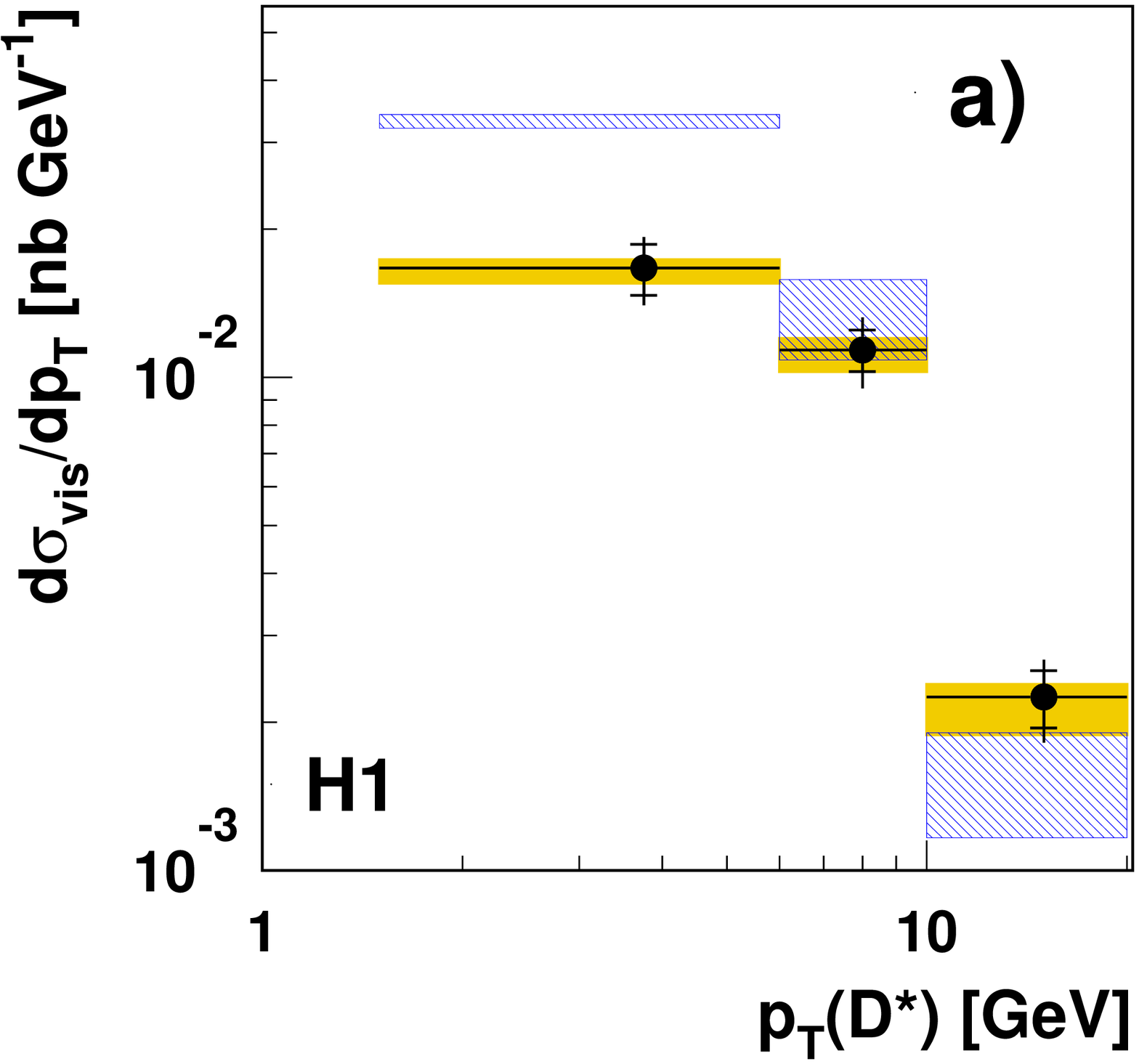}
\includegraphics*[width=6.5cm,height=6.4cm]{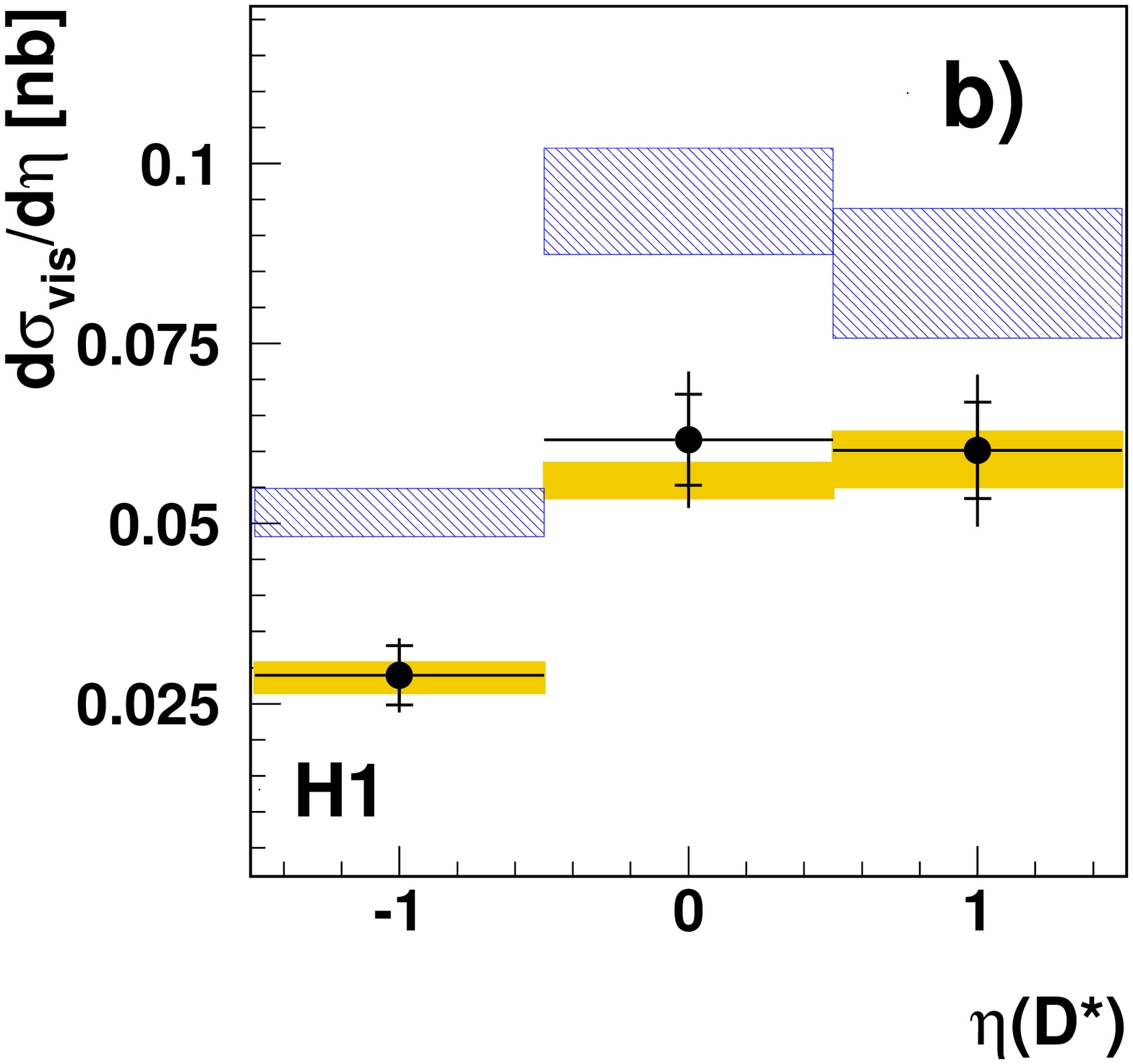}
\includegraphics*[width=6.5cm,height=6.4cm]{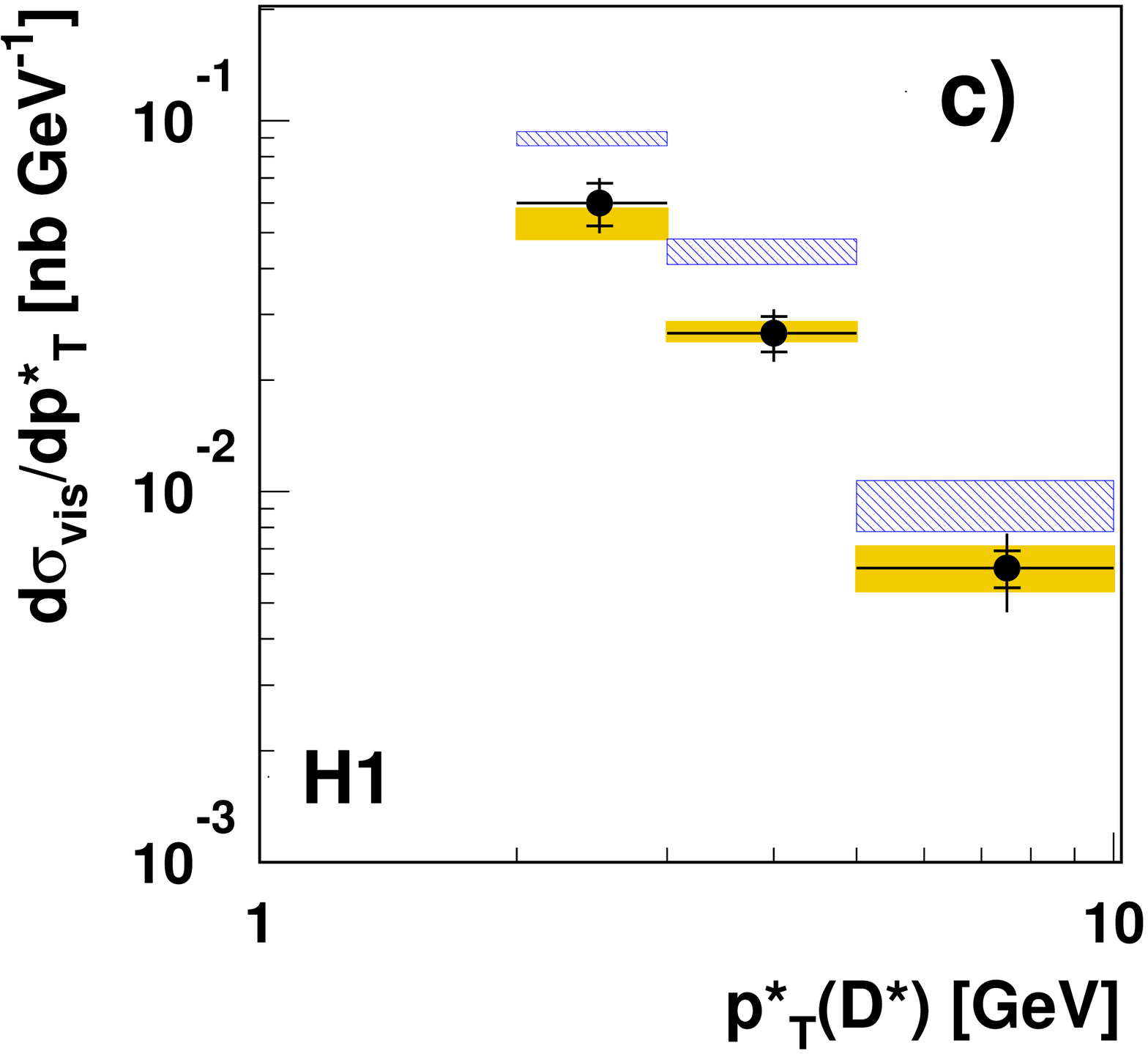}
\includegraphics*[width=6.5cm,height=6.4cm]{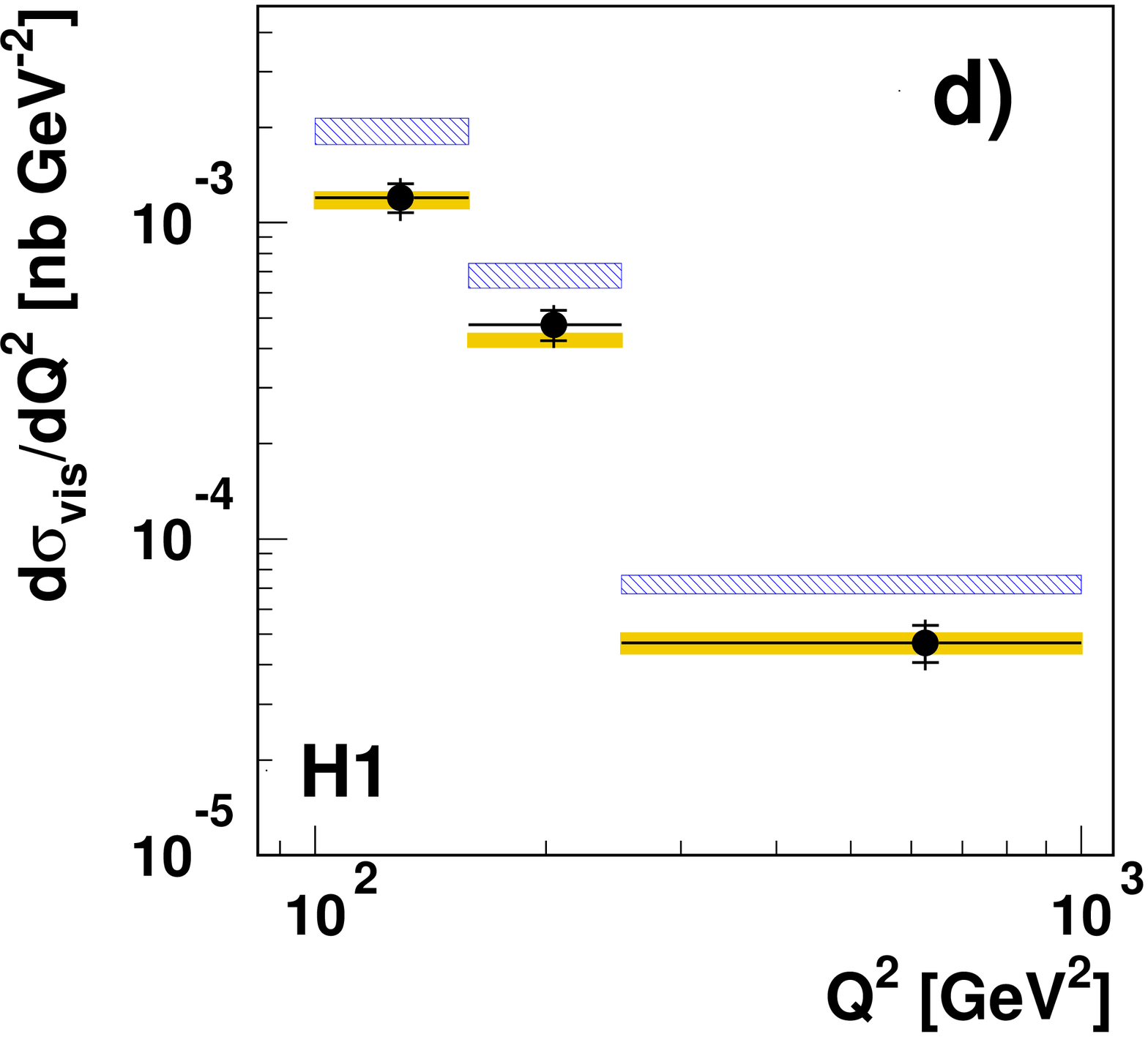}
\includegraphics*[width=6.5cm,height=6.4cm]{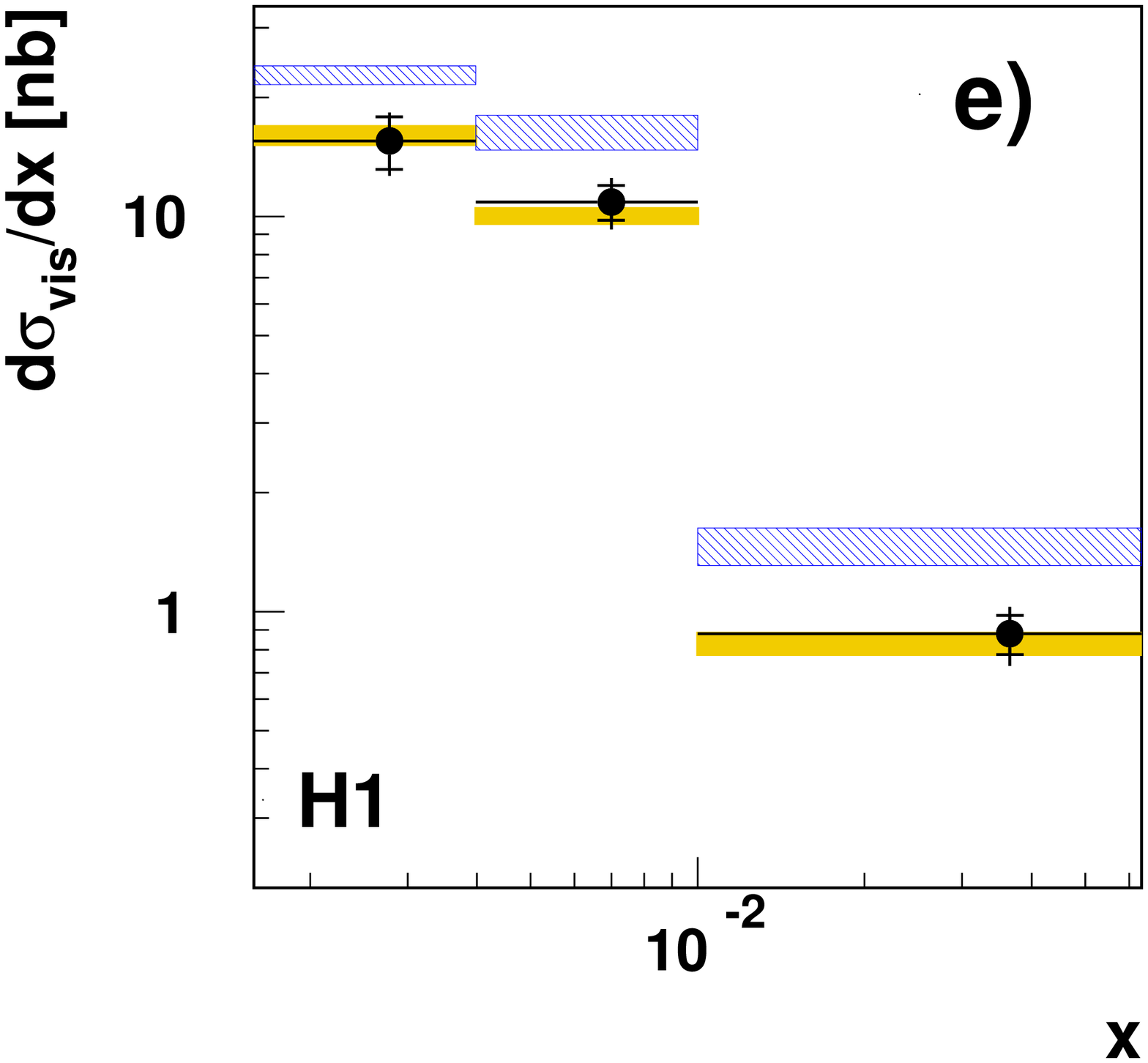}
\includegraphics*[width=6.5cm,height=6.4cm]{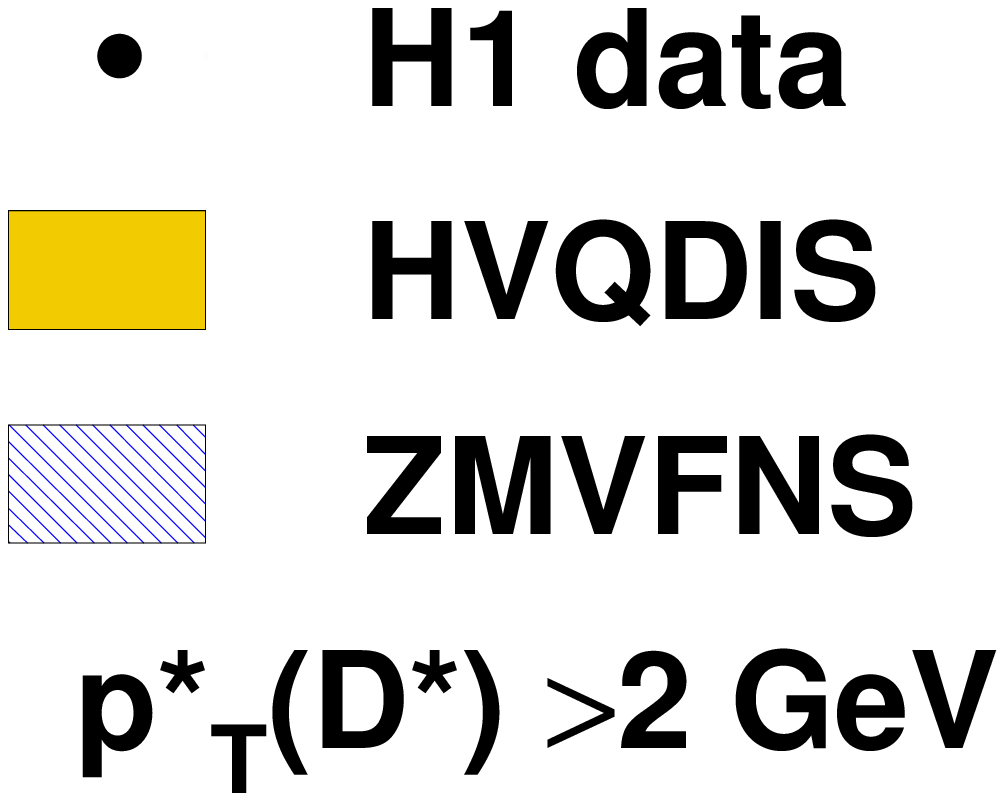}
\end{center}
\caption{Differential cross sections for inclusive \dstarpm meson production as a function of 
$p_T(D^*)$, $\eta (D^*)$, $p^*_T(D^*)$, $Q^2$ and $x$ as measured for $p^*_T(D^*)>2 \, \rm GeV$.
The inner error bars indicate the statistical uncertainties, the outer error bars show the statistical 
and systematic uncertainties added in quadrature. The expectation of HVQDIS (shaded band) is obtained 
using the parameter variation described in section~\ref{uncertainties}. The prediction in ZMVFNS is 
represented by the hatched band where the uncertainty originates from the scale variation.}

\label{fig:dsigma_nlomodels}
\end{figure}

The data are also compared to the ZMVFNS prediction~\cite{zmvfns}. This 
calculation has an intrinsic limitation on the transverse \dstarpm momentum in
the photon-proton center of mass frame, namely $p^*_T(D^*)>2\,{\rm GeV}$. 
Therefore the same additional cut is applied to the data and the cross section
is determined for the corresponding phase space. 
In Fig.~\ref{fig:dsigma_nlomodels} the \dstarpm cross sections are shown as a function of 
$p^*_T(D^*)$, $p_T(D^*)$, $\eta(D^*)$ and \qsq, together with the ZMVFNS and HVQDIS calculations. 
The ZMVFNS prediction fails to describe the data, while HVQDIS agrees well with the data.

\section{Extraction of \boldmath{\ftc}}
\label{ftwocharm}
The charm contribution \ftc$(x,Q^2)$ to the inclusive proton structure function \ft
is defined by the expression for the single photon exchange cross section 
for charm production:
\begin{equation}
\displaystyle
\frac{d^2\sigma^{c\bar{c}}}{dxdQ^2}=\frac{2\pi\alpha_{em}^2}{Q^4x}
\left( [1+\left(1-y\right)^2]\;F^{c\bar{c}}_2(x,Q^2)-y^2 F^{c\bar{c}}_L(x,Q^2)\right) \ \ ,
\end{equation}
where $\alpha_{em}$ is the electromagnetic coupling constant. Weak interaction effects are neglected.

The contribution from the structure function $F_L^{c\bar{c}}$ amounts to at most $3\%$~\cite{riemersma} 
in the present phase space and is neglected. The visible inclusive $D^{*\pm}$ cross sections $\sigma_{\rm{vis}}^{\rm{exp}}(y,Q^2)$ in bins of $y$ and $Q^2$ 
are converted to a bin centre corrected $F_2^{c\bar{c}}(\langle x\rangle,\langle Q^2\rangle)$ in the 
framework of a particular model using the relation:
\begin{equation}
F_2^{c\bar{c}}(\langle x\rangle,\langle Q^2\rangle)=
\frac{\displaystyle \sigma_{\rm{vis}}^{\rm{exp}}(y,Q^2)}
{\displaystyle \sigma_{\rm{vis}}^{\rm{theo}}(y,Q^2)}\cdot
F_2^{c\bar{c}~\rm{theo}}(\langle x\rangle,\langle Q^2\rangle) \ \ ,
\label{f2cexp}
\end{equation}  
where $\sigma_{\rm{vis}}^{\rm{theo}}$ and $F_2^{c\bar{c}~\rm{theo}}$ are the 
theoretical  predictions from the model under consideration. 
As in previous publications \cite{D_H1,D_ZEUS} the HVQDIS program and another 
program~\cite{riemersma} are used to calculate these quantities at NLO. CASCADE is not 
used for an \ftc\, extraction since it does not agree with the data (Fig.~\ref{fig:dsigma_xq}). 

The model uncertainties on the measurement of \ftc\, are estimated by varying the HVQDIS 
parameters as described in section~\ref{uncertainties}. The variations are made simultaneously 
in the calculation of the visible \dstarpm cross sections and in the 
prediction for \ftc. The total model uncertainties amount to $1-7\%$ and are dominated by 
the variation of the renormalisation and factorisation scales. 
The central values of \ftc\, with experimental and model uncertainties are
summarised in Table~\ref{Tab:f2charm}.
The fraction of the total \dstarpm cross section in the visible phase space, as predicted by HVQDIS and 
given by $\frac{\sigma(y,Q^2)^{theo}_{vis}}{\sigma(y,Q^2)^{theo}_{tot}}$, is also quoted and varies between
$0.4$ and $0.7$. 

\begin{figure}[htb]
\includegraphics*[width=16.5cm]{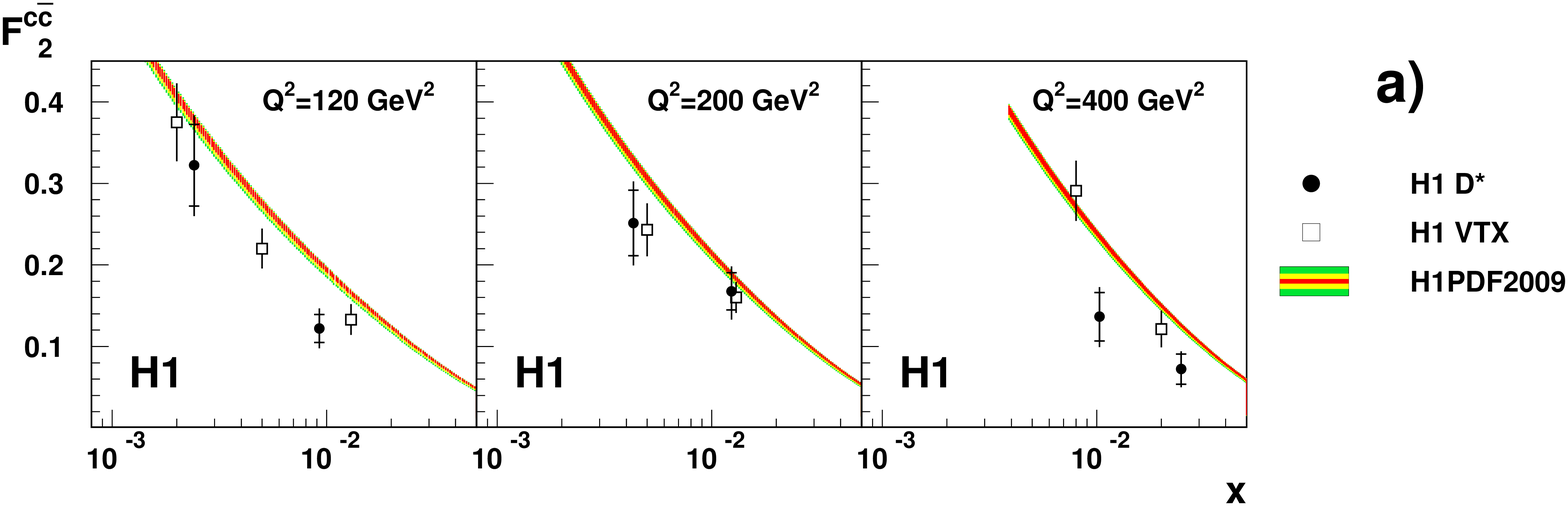}
\includegraphics*[width=16.5cm]{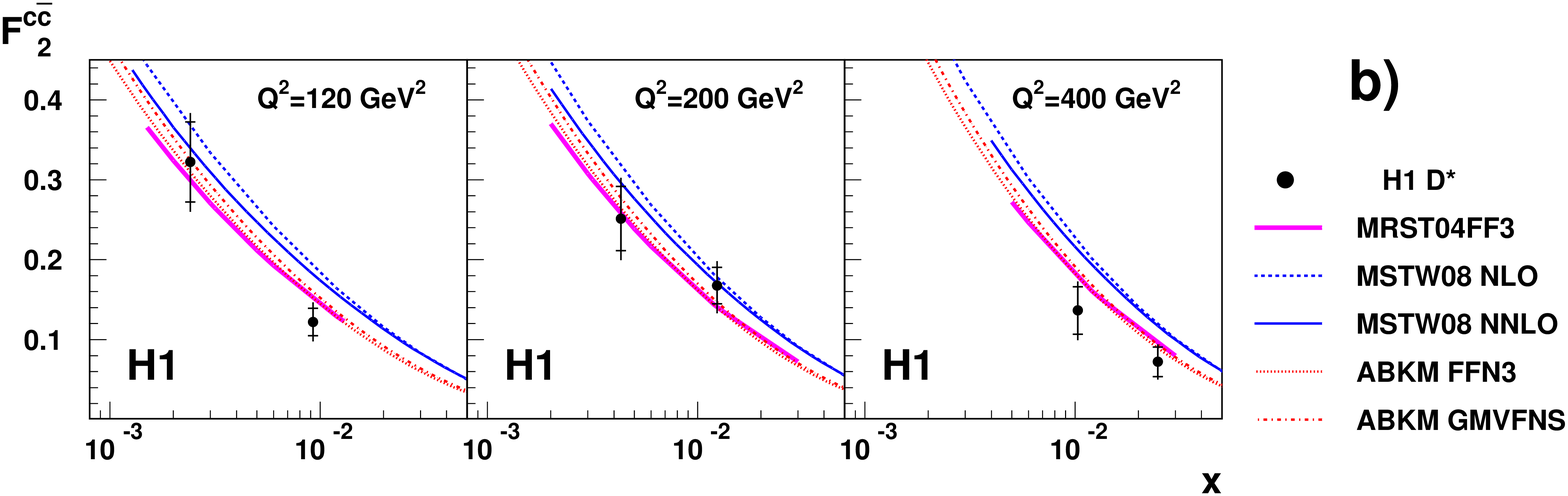}
\caption{The charm contribution \ftc\, to the proton structure function. The data (closed symbols) are 
shown with statistical (inner error bars) and total (full error bars) uncertainties. In a) the data 
are compared to the H1 measurement of \ftc\, using secondary vertex information (open symbols)~\cite{VTX_H1}, 
where measurements at \qsq$=300 \, \rm GeV^2$ are shifted to \qsq$=400 \, \rm GeV^2$ using the NLO 
calculation~\cite{riemersma}. The result of the PDF fit H1PDF2009 (shaded band) is also shown. The 
uncertainty band accounts for experimental, model and parametrisation uncertainties~\cite{h1f2bulk}.
In b) the data are compared to the QCD predictions from the NLO calculation~\cite{riemersma} in 
FFNS (light thick solid line). The predictions from the global PDF fits MSTW08 at NLO (dashed) and
NNLO (dark solid) as well as the results of the ABKM fit~\cite{abkm} at NNLO in FFNS (dotted) and 
GMVFNS (dashed-dotted) are also shown.}
\label{fig:f2charm}
\end{figure}

In Fig.~\ref{fig:f2charm} \ftc\, is shown as a function of $x$ for different values of $Q^2$. 
The \ftc\, values are consistent with those obtained in an inclusive track
measurement using the H1 vertex detector information~\cite{VTX_H1}.
The expectation from the recent PDF fit to 
inclusive DIS data, H1 PDF2009~\cite{h1f2bulk}, tends to overestimate the data.
In Fig.~\ref{fig:f2charm}(b) the measurements are compared to the massive FFNS 
calculation at NLO~\cite{riemersma} and NNLO~\cite{abkm} and to the 
GMVFNS predictions at NLO and NNLO~\cite{mstw,abkm}. 
The FFNS predictions agree well with the data over the full kinematic region investigated.
The expectations for \ftc\, from a global fit in the GMVFNS at NLO tend to overestimate the data.
At NNLO the GMVFNS prediction agrees better with the data. 

\section{Conclusions}

The cross section for \dstarpm meson production is measured in the phase space 
$100<Q^2<1000\,{\rm GeV}^2$ and $0.02 < y < 0.7$.
Single and double differential cross sections are compared to Monte Carlo 
simulations and the predictions of NLO calculations in massive and massless schemes.
The data have a typical precision of $20\%$.

In the measured domain the RAPGAP and CASCADE simulations do 
not provide good a description of the \dstarpm kinematics. The double-differential cross 
section $d^2\sigma/dy\, dQ^2$ is described well by CASCADE, while RAPGAP overestimates the 
cross section at high $Q^2$. The NLO FFNS calculation HVQDIS agrees with the data well, 
while the calculation based on ZMVFNS fails to describe the data. 

The charm contribution \ftc\, to the proton structure function \ft\, is determined. HVQDIS is used 
for extrapolation of the visible \dstarpm cross sections to the full phase space in $p_T(D^*)$ 
and $\eta(D^*)$. The model uncertainties are found to be 
small in the kinematic region studied. The data are compared to QCD predictions
at NLO in the FFNS scheme and to the CASCADE implementation of the CCFM model as well as 
to the expectations from global fit analyses, using 
GMFVNS implementations at NLO and NNLO. Both FFNS and CASCADE describe the measurement well. 
The data indicate that the NLO FFNS provides the best description of $D^*$ production 
and of \ftc\, in the kinematic region of the analysis.

\section*{Acknowledgements}

We are grateful to the HERA machine group whose outstanding efforts have
made this experiment possible. We thank the engineers and technicians for
their work in constructing and maintaining the H1 detector, our funding
agencies for financial support, the DESY technical staff for continual
assistance and the DESY directorate for support and the hospitality which
they extend to the non-DESY members of the collaboration. Furthermore we 
thank G. Kramer and C. Sandoval for fruitful discussions.

\newpage
\renewcommand{\arraystretch}{1.2}
\begin{table}
\centering
\begin{tabular}[t]{|c|c|c|c|c|}\hline
& & & &  \\[-0.3cm]
\rb{$p_T(D^*) \rm [GeV]$}   & \rb{$\frac{d\sigma}{dp_T} [\frac{\rm pb }{\rm GeV}]$} & \rb{$\delta_{stat}[\%]$} & \rb{$\delta_{unc}[\%]$} & \rb{$\delta_{cor}[\%]$}\\[-0.2cm] \hline
$1.5 \div 6.0$  & $27.8$ &  $11.1$ & $ 7.4$  & $^{+9.2}_{-8.9}$\\ 
$6.0 \div 9.5$  & $17.8$ &  $9.1 $&  $8.3$  & $^{+8.3}_{-8.4}$\\ 
$9.5 \div 20$   & $3.31$ &  $11.4 $&  $11.6$ & $^{+8.1}_{-8.1}$\\ \hline \hline 

& & & &  \\[-0.3cm]
\rb{$\eta(D^*)$}  & \rb{$\frac{d\sigma}{d\eta} [\rm pb]$}  & \rb{$\delta_{stat}[\%]$} & \rb{$\delta_{unc}[\%]$} & \rb{$\delta_{cor}[\%]$}\\[-0.2cm] \hline
$-1.5 \div -0.6$ & $51.5$ &  $12.0$ &  $7.5$ & $^{+7.4}_{-7.4}$\\ 
$-0.6 \div \phantom{-} 0.7$& $94.9$ &  $ 8.4$ &  $8.5$ & $^{+9.7}_{-9.6}$\\ 
$\phantom{-} 0.7 \div \phantom{-} 1.5$& $68.1$ &  $16.4$ &  $8.8$ &  $^{+8.0}_{-8.2}$\\ \hline \hline

& & & &  \\[-0.3cm]
\rb{$z(D^*)$} & \rb{$\frac{d\sigma}{dz} [\rm pb] $}  & \rb{$\delta_{stat}[\%]$} & \rb{$\delta_{unc}[\%]$} & \rb{$\delta_{cor}[\%]$}\\[-0.2cm] \hline
$0.0 \div 0.3$& $234$ &  $17.3$ & $ 7.8$ & $^{+8.9}_{-8.7}$\\ 
$0.3 \div 0.6$& $328$ &  $ 8.4$ & $ 8.3$ & $^{+8.6}_{-8.7}$\\ 
$0.6 \div 1.0$& $135$ &  $ 8.8$ & $ 9.0$ &  $^{+14.5}_{-13.8}$\\ \hline \hline

& & & &  \\[-0.3cm]
\rb{$\log(\frac{Q^2}{\rm GeV^2})$} & \rb{$\frac{d\sigma}{d \rm Q^2} [\frac{\rm pb}{\rm GeV^2}]$} & \rb{$\delta_{stat}[\%]$} & \rb{$\delta_{unc}[\%]$} & \rb{$\delta_{cor}[\%]$}\\[-0.2cm] \hline
$2.0 \div 2.2$& $1.88$ &  $10.1$ &  $7.6$ & $^{+8.6}_{-8.7}$\\ 
$2.2 \div 2.4$& $0.767$ &  $ 10.0$ & $ 8.2$ &  $^{+7.7}_{-7.6}$\\ 
$2.4 \div 3.0$& $0.0572$ &  $15.7$ & $ 9.6$ &  $^{+9.7}_{-9.7}$\\ \hline \hline

& & & &  \\[-0.3cm]
\rb{$\log(x)$} & \rb{$\frac{d\sigma}{dx} [\rm pb]$} & \rb{$\delta_{stat}[\%]$} & \rb{$\delta_{unc}[\%]$} & \rb{$\delta_{cor}[\%]$}\\[-0.2cm] \hline
$-2.8 \div -2.4$& $24.8 \times 10^3$ &  $13.2$ & $ 7.6$ &  $^{+6.9}_{-7.2}$\\ 
$-2.4 \div -2.0$& $16.0 \times 10^3$ &  $ 9.5$ & $ 8.0$ &  $^{+9.5}_{-9.1}$\\ 
$-2.0 \div -1.2$& $1.29 \times 10^3$ &  $12.3$  & $ 9.2$ &  $^{+10.2}_{-10.2}$\\ \hline
\end{tabular}
\caption{Single differential cross sections for \dstarpm production in bins of $Q^2$, 
$x$ and the meson kinematics, $p_T(D^*)$, $\eta(D^*)$ and $z(D^*)$, as measured in the 
visible range defined in  Table~\ref{tab:range}. The central values of the cross section are listed 
together with relative statistical ($\delta_{stat}$), uncorrelated 
($\delta_{uncor}$) and correlated ($\delta_{cor}$) systematic uncertainties.}
\label{Tab:dsigma}
\end{table}
\renewcommand{\arraystretch}{1.2}
\begin{table}
\small
\centering
\begin{tabular}[b]{|c|c|c|c|c|c|}\hline
& & & & & \\[-0.2cm]
\rb{$\log(\frac{Q^2}{\rm GeV^2})$} & \rb{$y$} & \rb{$\frac{d^2\sigma}{dQ^2 dy} 
[\frac{\rm pb}{\rm GeV^2}]$} & \rb{$\delta_{stat}[\%]$} &
\rb{$\delta_{uncorr}[\%]$} & \rb{$\delta_{corr}[\%]$} \\[-0.1cm] \hline
$2.0 \div 2.2$ &$ 0.020 \div 0.350$  &$ 3.39$ & $13.7$ &$ 7.6$ & $^{+11.6}_{-10.8}$\\ \hline
$2.0 \div 2.2$ &$ 0.350 \div 0.700$  &$ 2.11$ & $14.8$ &$ 7.6$ & $^{+6.4}_{-6.7}$\\ \hline
$2.2 \div 2.4$ &$ 0.020 \div 0.300$  &$ 1.61$ & $13.3$ &$ 8.2$ & $^{+8.0}_{-7.9}$\\ \hline
$2.2 \div 2.4$ &$ 0.300 \div 0.700$  &$ 0.810$ & $15.0$ &$ 8.2$ & $^{+7.6}_{-7.4}$\\ \hline
$2.4 \div 3.0$ &$ 0.020 \div 0.275$  &$ 0.0921$ & $24.8$ &$ 9.6$ & $^{+10.5}_{-10.4}$\\ \hline
$2.4 \div 3.0$ &$ 0.275 \div 0.700$  &$ 0.0803 $& $20.2$ &$ 9.6$ & $^{+9.7}_{-9.7}$\\ \hline
\end{tabular}
\caption{Double differential cross sections for \dstarpm production in bins of $Q^2$ and 
$y$ as measured in the visible range defined in Table~\ref{tab:range}. The central 
values of the cross section are listed together with relative statistical ($\delta_{stat}$), 
uncorrelated ($\delta_{unc}$) and correlated ($\delta_{cor}$) systematic uncertainties.}
\label{Tab:d2sigma}
\end{table}
\renewcommand{\arraystretch}{1.2}
\begin{table}
\small
\centering
\begin{tabular}[t]{|c|c|c|c|c|c|c|c|}\hline
& & & & & & & \\[-0.2cm]
\rb{$\langle Q^2\rangle [\rm GeV^2]$} & \rb{$\langle x \rangle$} &  \rb{\ftc} &
\rb{$\delta_{stat}[\%]$} & \rb{$\delta_{unc}[\%]$} & \rb{$\delta_{cor}[\%]$} &
\rb{$\delta_{model}[\%]$} & 
\rb{$\frac{\sigma(y,Q^2)^{theo}_{vis}}{\sigma(y,Q^2)^{theo}_{tot}}$} \\[-0.1cm] \hline
$120$ & $0.00924$ &$0.122$  &$ 13.7$ & $7.6$ & $^{+11.6}_{-10.8}$ &$^{+3.2}_{-3.8}$ & $0.53$\\ \hline
$120$ & $0.00241$ &$0.322$  &$ 14.8$ & $7.6$ & $^{+6.4}_{-6.7}$   &$^{+3.4}_{-4.8}$ & $0.63$\\ \hline
$200$ & $0.01240$ &$0.168$  &$ 13.3$ & $8.2$ & $^{+8.0}_{-7.9}$   &$^{+3.8}_{-4.6}$ & $0.48$\\ \hline
$200$ & $0.00432$ &$0.251$  &$ 15.0$ & $8.2$ & $^{+7.6}_{-7.4}$   &$^{+3.3}_{-3.5}$ & $0.67$\\ \hline
$400$ & $0.02480$ &$0.072$  &$ 24.8$ & $9.6$ & $^{+10.5}_{-10.4}$ &$^{+6.5}_{-5.9}$ & $0.43$\\ \hline
$400$ & $0.01030$ &$0.136$  &$ 20.2$ & $9.6$ & $^{+9.7}_{-9.7}$   &$^{+3.7}_{-3.8}$ & $0.71$\\ \hline
\end{tabular}
\caption{The measured values and relative errors for the charm contribution to the proton 
structure function \ftc. Relative statistical, correlated and uncorrelated experimental systematic as well
as model uncertainties are listed. The fractions of the total \dstarpm cross section 
in the visible phase space as predicted by HVQDIS are also given.}
\label{Tab:f2charm}
\end{table}
\renewcommand{\arraystretch}{1.0}

\newpage
\clearpage
\newpage
\newpage

\newpage

\newpage

\end{document}